\documentclass{aa}  
\usepackage{graphicx}
\usepackage{txfonts}
\usepackage{amsmath}
\usepackage[hidelinks]{hyperref}
\usepackage{url}
\usepackage{float}
\usepackage{afterpage}
\usepackage{amsmath, amssymb}
\usepackage{xcolor}
\usepackage{soul}
\usepackage{subcaption}
\usepackage{graphicx}

\begin{document} 

   \title{Follow-up of three exocomet-host candidates}

   \author{P. Muñoz-Cutanda
          \inst{1,2}
          \and
          I. Rebollido\inst{3}
          \and 
          B. Montesinos\inst{4}
          \and
          P. Cruz\inst{4}
          \and 
          O. Absil\inst{5}\fnmsep\thanks{F.R.S.-FNRS Research Director}
          \and
          S. Ertel\inst{6,7}
          }

\institute{Observatorio Astronómico Nacional (OAN-IGN), C/ Alfonso XII 3, 28014 Madrid, Spain\\
              \email{p.munoz@oan.es}
         \and
             Facultad de Ciencias Físicas, Pl. de Ciencias 1, Universidad Complutense de Madrid, 28040, Madrid, Spain
         \and
             European Space Agency (ESA), European Space Astronomy Centre (ESAC), Camino Bajo del Castillo s/n, 28692 Villanueva de la Cañada, Madrid, Spain
        \and
            Centro de Astrobiología (CAB), CSIC-INTA, ESAC Campus, Camino Bajo del Castillo s/n, 28692 Villanueva de la Cañada, Madrid, Spain
        \and
        STAR Institute, Université de Li\`ege, Allée du Six Août 19c, B-4000 Li\`ege, Belgium
        \and
        Department of Astronomy and Steward Observatory, University of Arizona, 933 N Cherry Ave., Tucson, AZ, 85721-0065, USA
        \and
         Large Binocular Telescope Observatory, University of Arizona, 933 N Cherry Ave., Tucson, AZ 85721-0065, USA
             }

   \date{Received October 28, 2025; accepted January 19, 2026}

  \abstract
   {Exocomets are small bodies that evaporate when they approach their host star. They are detected through variability of non-photospheric features with spectroscopy and/or asymmetric transits in time-series photometry. In the past four decades $\sim$30 systems have shown such variations, and were therefore classified as exocomet host stars. However, some publications have pointed out mechanisms that might mimic exocometary features, and therefore, careful monitoring is needed to confirm the origin of the observed variability.}
   {With this paper we aim to investigate the exocomet nature of the non-photospheric variable features observed in the exocomet candidate stars HD~36546, HD~42111 and 
   HD~85905. All of them have shown some degree of variability, particularly in their Ca {\sc ii} K line.}
   {We analysed the non-photospheric Ca {\sc ii} K line features from high-resolution spectra obtained using new NOT/FIES and Mercator/HERMES, and some additional archival spectra of the target stars. The variability was quantified through the changes in the equivalent widths of those features, which are assumed to be of circumstellar origin. Column densities were also estimated for each variable feature.}
   {Strong variability was found for HD~85905, consistent with a potential link to exocometary activity. However, the binarity of the system, which we confirmed through interferometric VLTI/PIONIER observations, complicates the interpretation of these signatures and prevents us from drawing definitive conclusions. The remaining two sources do not show any significant variability, but due to the sporadic nature of the exocometary events, we cannot discard the exocomet hypothesis. Further monitoring of the stars will be necessary to carry out a robust determination of the variability patterns and timescales that would completely rule out other scenarios.}
   {}

   \keywords{(Stars:) circumstellar matter; Stars: individual: HD~36546; Stars: individual: HD~42111; Stars: individual: HD~85905; Comets: general}

   \maketitle

\section{Introduction}
\label{sec:intro}
Since the discovery of exocomets around the $\beta$ Pic system \citep{Hobbs85,Ferlet87}
the interest in the study of minor bodies around other stars, in spite of the inherent difficulties for their detection, has increased. Comets and minor bodies are the most abundant objects in the solar system, some works pointing to the fact that the Oort cloud can host of the order of $10^{12}$ comets \citep{Dones04}; on the other hand, their potential contributions to supplying water or organic molecules associated to life on Earth are also relevant \citep{Lis19,Val-Borro14,Cao24}. In addition to the intensive search for exoplanets, the detection of exocomets is an important piece of the puzzle to understand the processes involved in the formation of planetary systems.

Exocomets are small bodies that evaporate as they approach their host star, developing a trail of dust and/or gas, that resembles the coma and tail of comets in the solar system. They can be detected through the variability in the circumstellar absorption components superimposed on the photospheric Ca {\sc ii} K line \citep[e.g.][]{Kiefer14b}, or in their photometric light curve as the exocomets transit before the star \citep{Zieba19,Pavlenko22,Dumond25,Norazman25}. The spectroscopic variability of the non-photospheric features is interpreted within the so-called ``falling evaporating bodies'' (FEBs) scenario \citep{Beust90}, in which the icy bodies sublimate as they approach the star, releasing a gas cloud that produces the circumstellar absorptions. Around $\sim\!30$ systems are known to exhibit features in their spectra that could be attributed to exocomets \citep{Rebollido20,Strom20}. Most of the objects are rapidly rotating A-type stars. Only a few systems, including some F-type stars, show photometric variations, with $\beta$ Pic \citep{Lecavelier22} and HD~172555 \citep{Kiefer23} being the only ones where exocomets have been found both in spectroscopy and photometry. However, spectral variations are not always produced by exocomets: some works \citep[e.g.][]{Montesinos19,Eiroa21} have shown that long-term monitorings over different 
timescales are necessary to discard different alternative scenarios (binary systems, circumstellar disks or envelopes). 

The importance of debris disks cannot be ignored in this context, since they are linked to exocometary activity \citep{Rebollido20}. Debris disks are made up of dust and, to a lesser extent, gas, although historically they were thought to be gas free because the host star ultraviolet photons can photodissociate the gas in protoplanetary disks in short timescales. However, CO has been found in $\sim$25 systems \citep[e.g.][]{Marino16,Kral20,Rebollido22,Moor25} in far-infrared and sub-millimetre observations, and measurements in CO absorption lines can set up a limit to the vertical distribution of gas in debris disks \citep{Worthen24a}. Whether gas in debris disks is a remnant of protoplanetary disks or second-generation material is still a debate.

In this work we have analysed high-resolution spectra of three stars with significant non-photospheric variations in the Ca {\sc ii} K line, that were attributed to exocomets. The paper is organised as follows: Sect. \ref{sec:stars} summarizes some properties of the three stars, in Sect. \ref{sec:observations} details of the observations, telescopes and instruments, setups and dates are given, Sect. \ref{sec:methods} describes the methodology used, Sect. \ref{sec:results} and \ref{sec:discussion} contain the results and the discussion, making special emphasis on HD~85905, given its newly discovered binary architecture and the large variations observed, and in Sect. \ref{sec:conclusions} some conclusions are drawn.

\section{Selected sample}
\label{sec:stars}

The three objects selected for this study have shown non-photospheric variable features in the Ca {\sc ii} K line \citep{Rebollido20}. Table \ref{tab:parameters} shows some of their basic stellar parameters. In order to put this work in context, in what follows we give some specific details of the objects studied:

\begin{table}[t]
    \caption{Stellar parameters}
    \label{tab:parameters}
    \centering
    \setlength{\tabcolsep}{2.5 pt}
    \begin{tabular}{ccccccc}
    \hline\hline
    \noalign{\smallskip}
    HD & SpT& $T_{\rm eff}$ & $\log g$ & $\varv\;\sin i$ & [M/H] & v$_{\rm rad}$ \\
         &        &     (K)       &          &   (km/s)       &      &   (km/s) \\ 
    \noalign{\smallskip}
    \hline
    \noalign{\smallskip}
    36546 & B8 V  & 9150$\pm$50 & 4.44$\pm$0.10 & 150 & $-1.0$ & 14.7$\pm$0.6\\
    42111 & A3 V  & 9380$\pm$40 & 3.48$\pm$0.10 & 252 &  0.0 & 27.5$\pm$1.9\\
    \noalign{\smallskip}
    \hline
    \noalign{\smallskip}
    85905 & A1 IV & \multicolumn{4}{c}{See Section \ref{sec:hd85905}}\\
    \noalign{\smallskip}
   \hline        
\end{tabular}
\tablefoot{The results for HD~36546 and HD~42111 are from \cite{Rebollido20}. 
The parameters for HD~85905 were also determined in that paper under the assumption that the star was a single object. They are reassessed in this work after the discovery that the star is actually a binary. To avoid 
confusion, the single-object parameters by \cite{Rebollido20} are not listed
here and we defer the reader to Sect. \ref{sec:hd85905} and Appendix \ref{app:photometryandhr} for details on the estimation of the parameters of the binary components.}
\end{table}

HD~36546: This young star hosts a cold debris disk that was first imaged by \cite{Currie17} using Subaru/SCExAO, with a subsequent multi-wavelength study carried out by \cite{Lawson21}. The debris disk around the star is not observed perfectly edge-on ($i\!\sim\!70^\circ\!-\!75^\circ$). With an age estimated to be $\sim$3-10 Myr, the system is an ideal target to study a debris disk shortly after its formation. \cite{Lisse17} concluded that the disk had already lost its primary gaseous composition, and reported an overabundance of C that could be linked to a recent giant collision. Furthermore, using ALMA observations, $^{12}$CO and $^{13}$CO gas in the disk and in the continuum was detected \citep{Rebollido22}.

HD~42111: \cite{Grady96} reported a weak shell absorption using IUE (International Ultraviolet Explorer) high-dispersion spectra with the strongest absorption centred on the star radial velocity, pointing to the 
presence of low-velocity circumstellar gas around the star. \cite{Welsh13} studied the circumstellar environment and detected weak absorption features superimposed on the Ca {\sc ii} K line and a sporadic FEB event. 

HD~85905: Sporadic circumstellar Ca {\sc ii} and Na {\sc i} absorptions, similar to those observed in $\beta$ Pic, were interpreted within the FEB scenario \citep{Welsh98,Redfield07}. In the latter work, the significant Ca {\sc ii} column densities estimated and the variability observed could not be explained by interstellar medium (ISM) absorptions, which led the authors to propose the existence of a dust and gas disk around the star. In the present work, we use VLTI/PIONIER observations to show that the star is an almost-equal flux binary (see Sect.~\ref{sec:hd85905}), creating an even more complex scenario to explain its variability.

The variability reported for the three objects in \cite{Rebollido20} was spotted from observations carried out in the period 2015-2017 with the same telescopes and instruments detailed in Sect. \ref{sec:observations} used for this work \citep[see Table 1 of][for details]{Rebollido20}. We defer the reader to section 4.2 of that paper
and the associated figures showing comparisons among spectra of
the different campaigns (figures 8, 9 for HD~36546, 13, 14 
for HD~42111, and 17, 18 for HD~85905). The kind of variability observed was of different type: a typical red-shifted $\beta$ Pic-like
event in HD~36546, subtle changes of the Ca {\sc ii} H and K 
profiles in HD~42111, and dramatic variations both in Ca {\sc ii} K and
Na {\sc i} D in HD~85905. These phenomena were the motivation of 
further monitoring campaigns, carried out in the period 2022--2025,
whose results are presented in this paper.

\section{Observations}
\label{sec:observations}

\subsection{Optical spectroscopy}

The observations were obtained in different campaigns from 2012 to 2025 using the FIES (FIbred-fed Echelle Spectrograph) and HERMES (High Efficiency and Resolution Mercator Echelle Spectrograph) instruments on the Nordic Optical Telescope (NOT) and Mercator telescope, respectively, both located at El Roque de los Muchachos Observatory in La Palma, Spain, and FEROS (Fiber-fed Extended Range Optical Spectrograph) on the MPIA 2.2-m telescope at La Silla Observatory, Chile. The resolutions and wavelength ranges covered by each spectrograph are:  $R\!\sim\!67\,000$, $\Delta\lambda$=3700 -- 8300 \AA{ }(FIES), $\sim\! 85\,000$, 3770 -- 9000 \AA{ } (HERMES), and $\sim\!48\,000$, 3500 -- 9200 \AA{ } (FEROS). Two archival spectra, obtained with Keck/HIRES (2012, see Table \ref{tab:alldata} and Fig. \ref{HD85905all}) were also added to our own observations and used in the analysis. 

The spectra were reduced using the instruments' pipelines and corrected from heliocentric velocity. Since this work focuses on the analysis of the circumstellar absorptions superimposed on the photospheric Ca {\sc ii} K line at 3933.66 \AA, the removal of telluric lines was not necessary. The mean signal-to-noise ratios (S/N), taking into account all the spectra for each target, are 118 (HD~36546), 121 (HD~42111), and 95 (HD~85905). For the case of 
HD~85905 individual spectra with S/N$<$70 were not used in the analysis due to the complexity of the Ca {\sc ii} K line (see Sect. \ref{sec:results}). However, in the first step described in Sect. \ref{sec:methods}, where a weighted mean spectrum is obtained, all available spectra were included. This selection criterion was not required for HD~36546 and HD~42111, as a smaller number of features were detected. 

Table \ref{tab:alldata} and Fig. \ref{HD85905all} provide --in addition to quantitative results that will be discussed later-- a complete summary of
the number of spectra obtained for each star, dates and instruments used. 

\begin{figure*}[!ht]
    \centering
    \includegraphics[height=14cm]{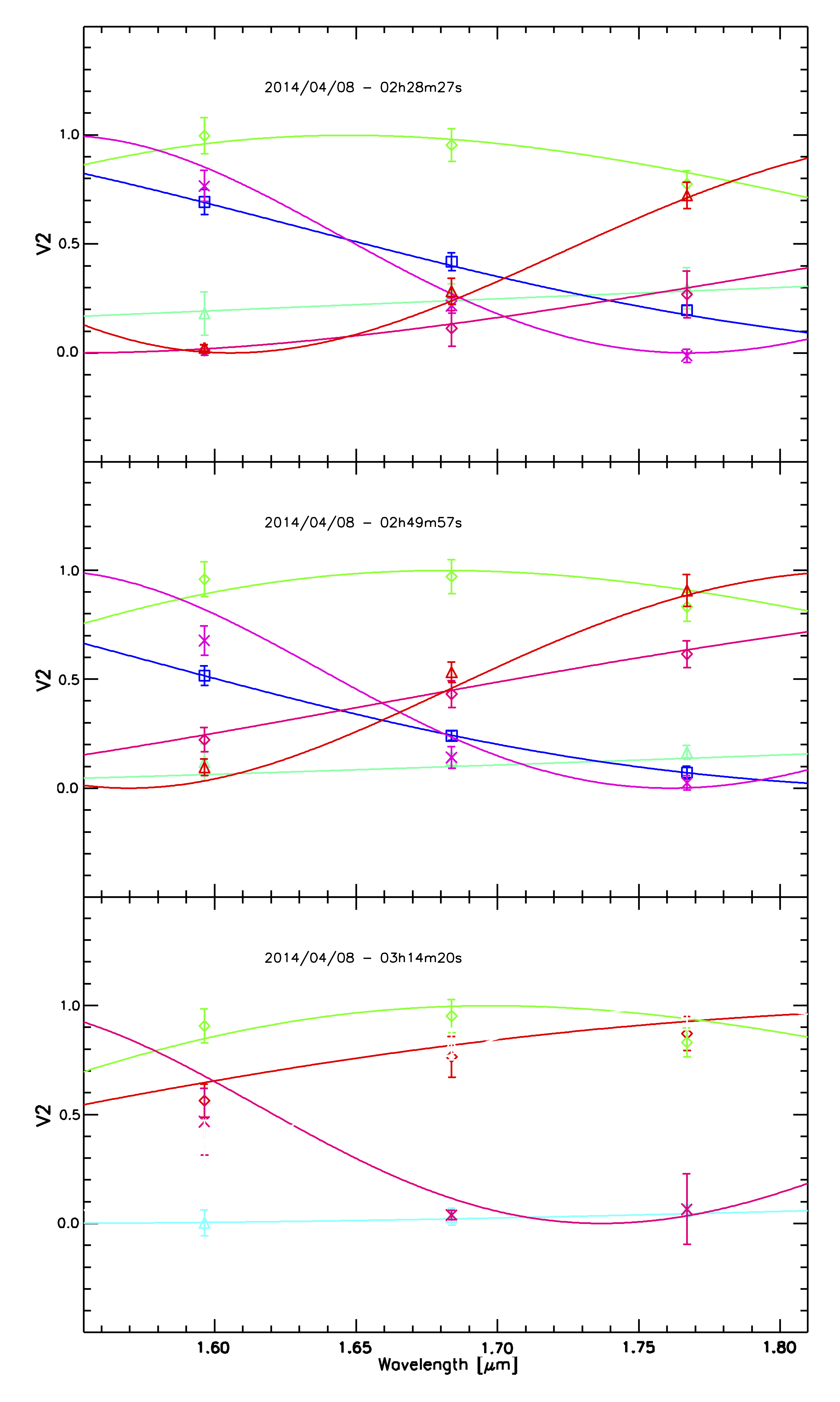}
    \includegraphics[height=14cm]{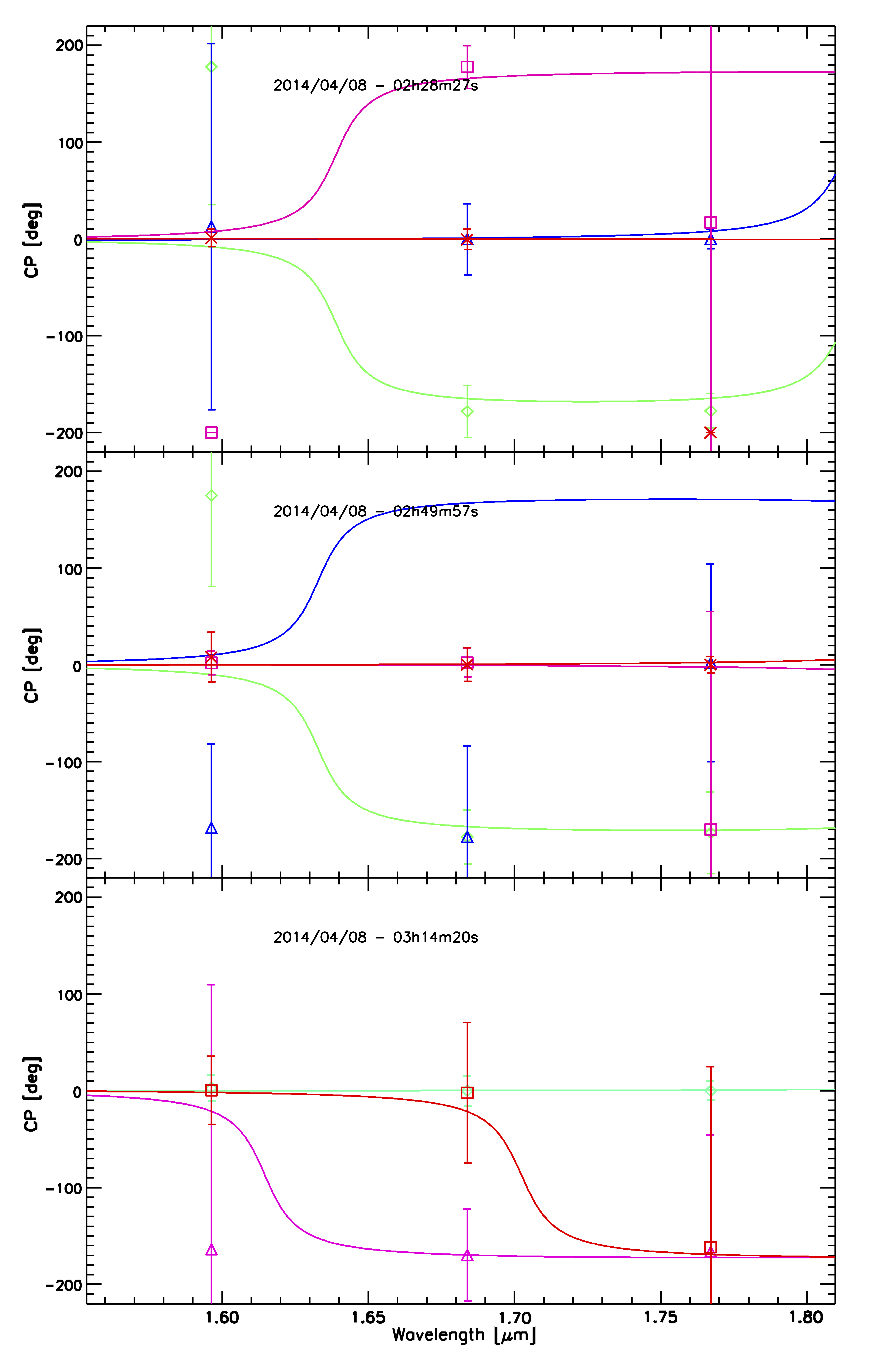}
    \caption{\textit{Left.} Squared visibilities as a function of wavelength, obtained for the three observing blocks (three panels) and six baselines (depicted with six different symbols and colours), along with the best-fit binary model described in Sect.~\ref{sec:binarity} (solid lines). \textit{Right.} Same as left, for the four closure phases obtained in each observing block. Some of the squared visibilities and closure phases are not displayed for the third observing block, and not used in this analysis, due to lower data quality.}
    \label{fig:pionier_data}
\end{figure*}

\subsection{Infrared interferometry}

HD~85905 was observed on 8 April 2014 with the VLTI/PIONIER interferometer as part of the exozodiacal disk survey of \citet{Absil21}. The four 1.8-m Auxiliary Telescopes (ATs) were used in the D0-H0-G1-I1 array configuration featuring baselines ranging between 41~m and 82~m, producing six visibility and four closure phase measurements simultaneously. The PIONIER read-out mode was set to FOWLER with SMALL dispersion (three spectral channels). Four calibrator stars were selected from \citet{Merand05} within 10$^\circ$ on the sky to minimize the effects of pupil rotation or instrumental polarization \citep[see][]{Lebouquin12}. Additional selection criteria were an $H$-band magnitude similar to the science target, and a small angular diameter, resulting in the following four calibrator stars: HD~83844 (CAL1), HD~84639 (CAL2), HD~86391 (CAL3), and HD~89338 (CAL4). All four are of K0 III spectral type, with $H$-band magnitudes ranging from 5.4 to 5.9. The target was observed in a CAL1-SCI-CAL2-SCI-CAL3-SCI-CAL4 sequence, taking a total of about 1h 15m, under variable, yet decent atmospheric conditions: seeing ranging from 0\farcs7 to 1\farcs2, coherence time around 2.5~ms. The data reduction consists of the conversion of raw observations into calibrated interferometric observables (squared visibilities and closure phases). We use the exact same method as in \citet{Ertel14}, where we calibrate each SCI of the CAL-SCI-...-CAL sequence individually by pairing it with either the preceding or the following CAL. The extracted squared visibilities and closure phases are displayed in Fig.~\ref{fig:pionier_data}.

While this observation was part of a larger survey, this data set was not included in the related publication \citep{Absil21} because HD~85905 was immediately identified as a binary deserving a dedicated follow-up study, based on the strong visibility variations and closure phases observed in the PIONIER data. As part of our follow-up attempt, the target was observed again with PIONIER on 1 February 2018, but only one observing block was obtained in cloudy conditions. This data set, of lesser quality, does not bring useful information in the context of the present study (not enough for a proper orbital characterization) and is not further considered here.

\section{Methodology}
\label{sec:methods}

\begin{figure*}[!ht]
    \centering
    \includegraphics[width=1.0\linewidth]{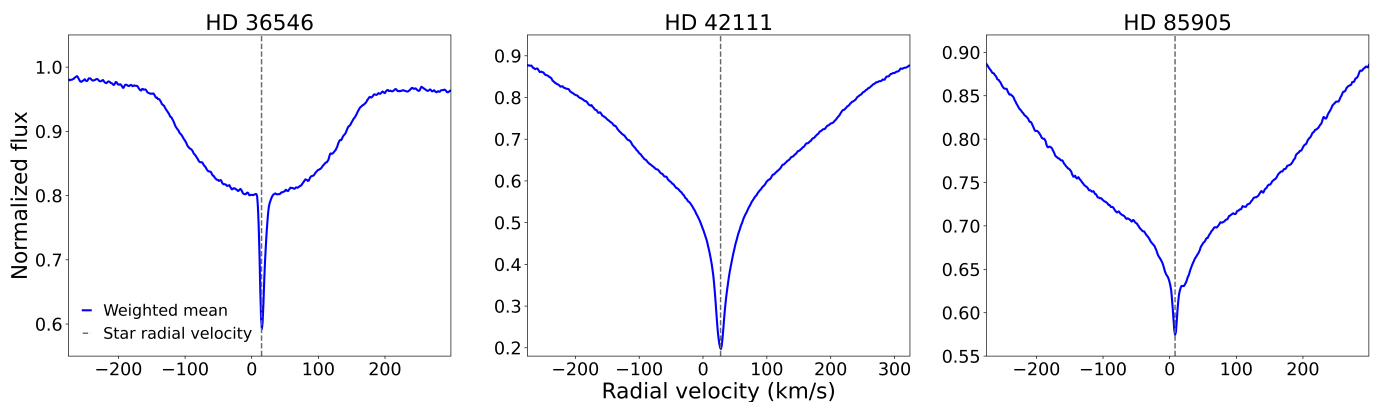}
    \caption{Weighted mean profiles of the Ca {\sc ii} K absorption line of HD~36546, HD~42111 and HD~85905 using all the available spectra for each object. The circumstellar non-photospheric narrow components are very distinguishable from the photospheric absorptions.}
    \label{meanspectra}
\end{figure*}

In order to search for and analyse any possible variations in the circumstellar components, weighted means of all the spectra were obtained by taking into account their S/N in the region of the Ca {\sc ii} K line. The S/N is defined as $\mu/\sigma$, where $\mu$ is the arithmetic mean of the flux and $\sigma$ is the standard deviation, both calculated in a 1-2 \AA~interval in the continuum close to the line. The weighted mean, $\bar{x}$, is computed using the expression:
\begin{equation}
\label{eq:mean}
    \bar{x}=\frac{
    \sum\limits_{i=1}^{n} \omega_i x_i
    }{
    \sum\limits_{i=1}^{n} \omega_i},
\end{equation}
\noindent  where $\omega_i=1/\sigma_i^2$ is the weight of the $i$-th spectrum. Fig. \ref{meanspectra} shows the weighted means of the Ca {\sc ii} K profiles computed using all the observations available for each object. The photospheric profile of $\bar{x}$ for each star was fitted to a spline; each individual Ca {\sc ii} K profile was divided by the spline, the results being the circumstellar non-photospheric contributions normalized
to intensity 1.0. Equivalent widths (EW), and the uncertainties ($\delta$EW), of each circumstellar profile were computed using the expressions:
\begin{equation}
\label{eq:ew}
    \mathrm{EW}=\int_{\lambda_1}^{\lambda_2}\frac{F_\mathrm{c}-F_\lambda}{F_\mathrm{c}} \;\mathrm{d}\lambda=\int_{\lambda_1}^{\lambda_2}\left(1-\frac{F_\lambda}{F_\mathrm{c}}\right)\;\mathrm{d}\lambda
\end{equation}

\noindent where $F_\lambda$ is the line flux, $F_\mathrm{c}$ is the flux of the continuum, and the integration interval $[\lambda_1, \lambda_2$] ($\Delta\lambda=\lambda_2-\lambda_1$) brackets the non-photospheric component. 
The EWs of the circumstellar components were computed after dividing the full Ca {\sc ii} K profile of by a spline to get rid of the 
photospheric contribution of the line. The process is illustrated in Fig. \ref{spline}. The uncertainties in the EWs have been computed 
using the comprehensive formalism described in Appendix 1 of the paper by \cite{Howarth86} and implemented in the Starlink\footnote{\url{https://starlink.eao.hawaii.edu/starlink/WelcomePage}} 
software {\sc dipso} which has been used in this work, and in particular
for this task. The results for the EWs and the $2\sigma$ uncertainties are given in Table \ref{tab:alldata} and a graphical display of these data is provided in Fig. \ref{fig:EW_time_ev}.

\begin{figure}[!ht]
    \centering
    \begin{subfigure}{1.02\linewidth}
        \centering
        \includegraphics[width=1\linewidth]{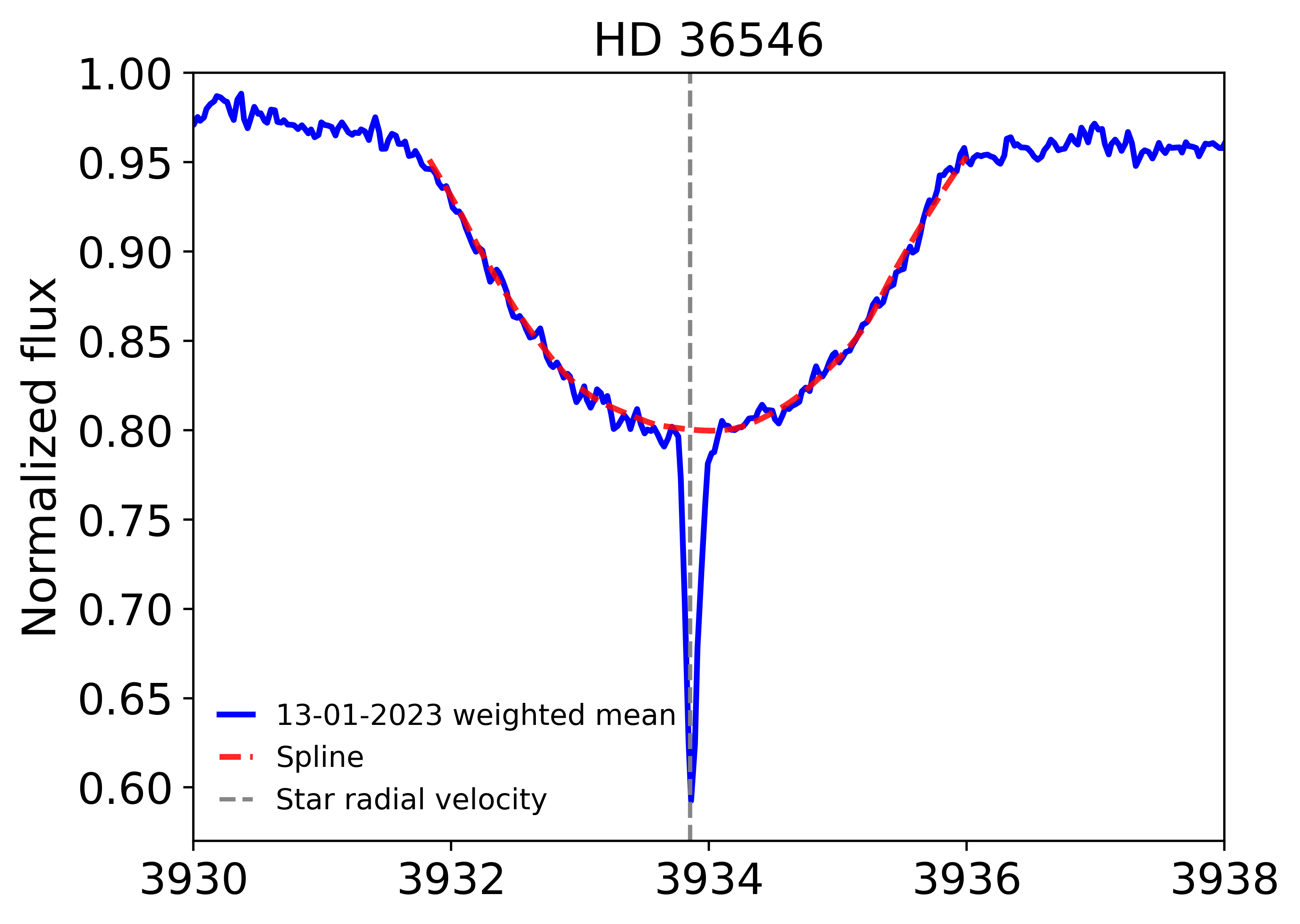}
        \label{splinea}
    \end{subfigure}
    \vspace{0cm}
    \begin{subfigure}{1 \linewidth}
        \centering
        \includegraphics[width=1\linewidth]{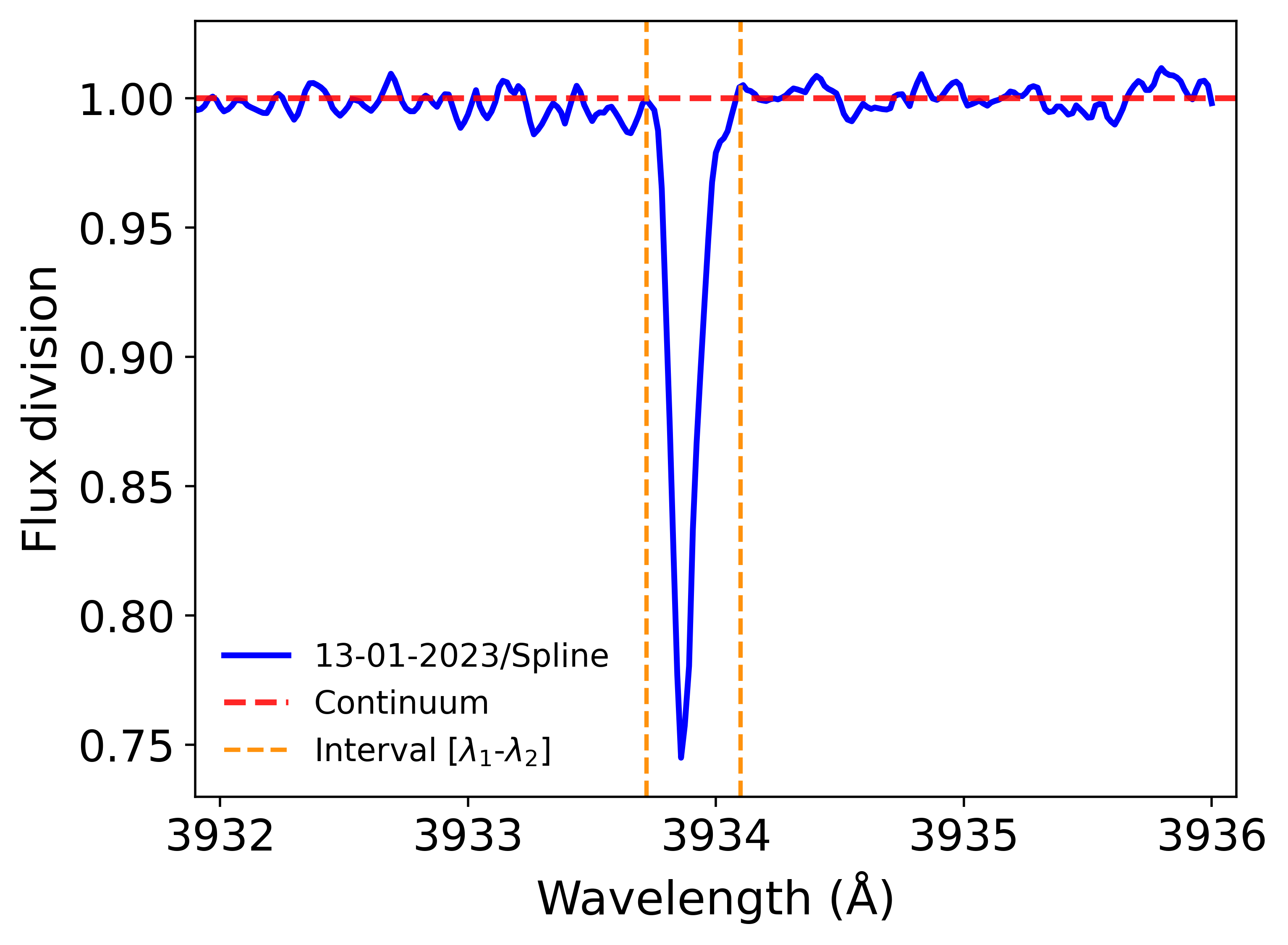}
        \label{splineb}
    \end{subfigure}
    \caption{\textit{Upper panel}: Weighted mean of the spectra of HD~36546, obtained on 13 Jan 2023, around the Ca {\sc ii} K line (blue), and the spline fitting the photospheric profile of the weighted mean spectrum built from all the available observations (red dashed line). The vertical grey dashed line marks the radial velocity of the star. \textit{Lower panel}: Ratio between the weighted mean spectrum from that date and the spline. The two vertical orange dashed lines mark the limits of the interval where the non-photospheric absorption is detected.}
    \label{spline}
\end{figure}

An estimate of the column density, N(Ca {\sc ii}), according to equation (1) in \cite{Somerville88} was computed:
\begin{equation}
\label{eq:somerville}
    {\rm N}({\rm Ca\,\,\textsc{ii}})=1.130\cdot10^{20}\,{\rm EW}/f_{12}\,\lambda_0^2\,\,\,{\rm cm}^{-2},
\end{equation}
\noindent where EW and $\lambda_0$ are the equivalent width and wavelength of
the line (both of them in \AA), and $f_{12}$ is the oscillator strength of the transition; according to the NIST\footnote{\url{https://physics.nist.gov/PhysRefData/ASD/lines_form.html}} database, $f_{12}=0.682$. This expression is only applicable to optically thin lines.

The Ca {\sc ii} column density could be translated into total mass by making some strong assumptions about the local chemistry, geometry, density distribution of the gas, etc. The steps to carry out that estimate would be: (i) convert the column density $N$(Ca {\sc ii}) into $N$(Ca), (ii) compute the H abundance from the Ca abundance, so the hydrogen column density can be obtained, (iii) estimate the surface mass density, (iv) adopt a geometry, and (v) assume a ratio gas/dust. The main caveats, which
would lead to large uncertainties would be: (a) the optical depth of the absorption,
since eqn. \ref{eq:somerville} is only valid in the optical thin case, (b) the 
ionization fraction $\varepsilon$(Ca {\sc ii})/$\varepsilon$(Ca) -where $\varepsilon$ is the number particles-, which is unknown,
(c) the composition [Ca/H], that could be different from the stellar one, (d) the
geometry or covering factor if the absorbing gas is clumpy or intercepts a small fraction
of the star, (e) the gas-to-dust ratio, which is usually taken as 100 for 
the ISM or protoplanetary disks, could not apply to the case studied here. Given all these unknowns, we just provide the reader with the values of the Ca {\sc ii} column densities in Table \ref{tab:alldata}, so all variables involved can be explored and fixed according to the chosen hypotheses and scenarios.  

\section{Results}
\label{sec:results}

\subsection{General considerations}

In the three stars studied a stable circumstellar component is seen in all the spectra. These components do not vary their radial velocity, and they are particularly conspicuous and narrow in HD~36546 and HD~85905, and somewhat broader in HD~42111 (see Fig. \ref{meanspectra}). These components are located at the corresponding radial velocity of each star and therefore we assume they arise from a circumstellar disk or envelope. As it can be seen in Fig. \ref{meanspectra}, the weighted means of the circumstellar absorptions of HD~36546 and HD~85905 both show asymmetries in the red wing, more noticeable in the latter.

\begin{figure}[!ht]
    \centering
    \includegraphics[width=1.0\linewidth]{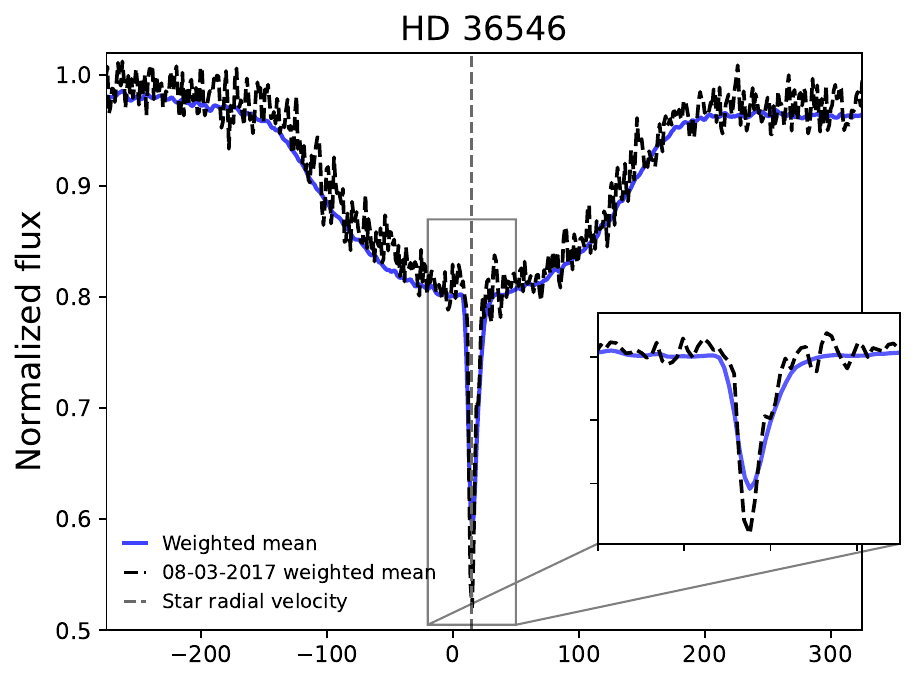}
    \includegraphics[width=1.0\linewidth]{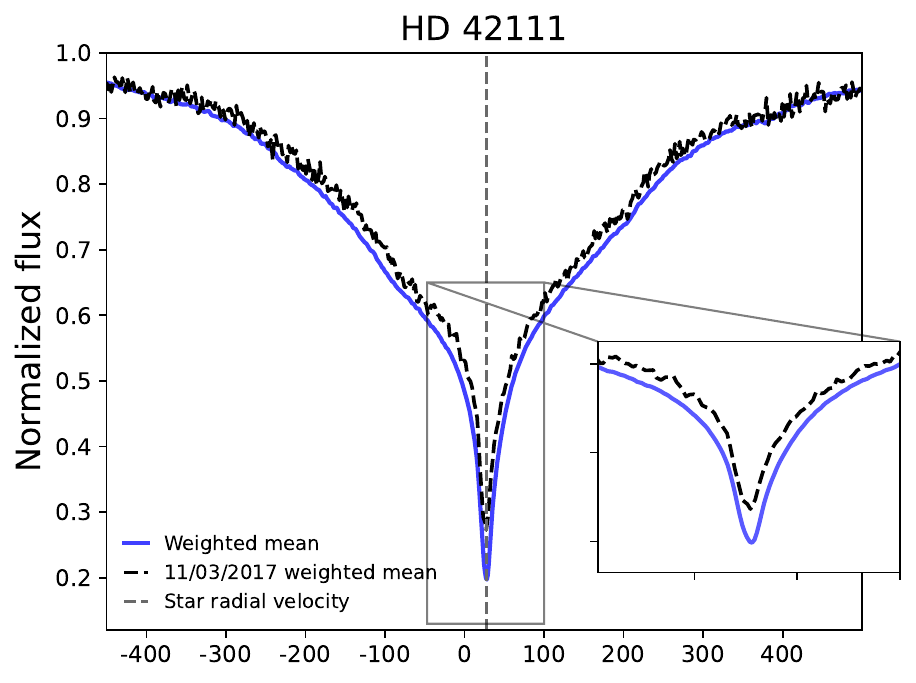}
    \includegraphics[width=1.0\linewidth]{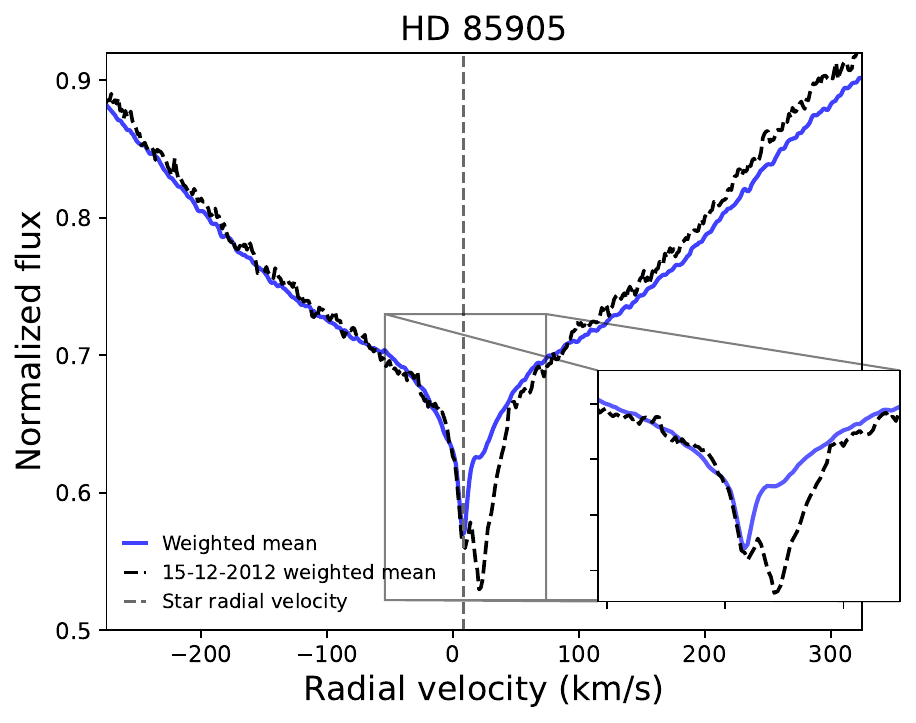}
    \caption{Maximum variation in the circumstellar line profiles of the stars with respect to their corresponding mean spectrum. For HD~36546 and HD~42111, the insets in both panels show tiny differences between the circumstellar features. For HD~85905, there is a remarkable feature around 25 km/s, similar to those observed in \cite{Beust90}, making it compatible with the FEB scenario.}
    \label{maxvars}
\end{figure}

Table \ref{tab:paramsfromews} shows, in the top half, the values 
of the equivalent widths of the circumstellar components of the Ca {\sc ii} K line obtained from the weighted average spectrum built for each object from the whole collection of individual spectra, and the corresponding values of the column densities estimated from these equivalent widths. Flux variations
of different degrees have been detected in the circumstellar Ca {\sc ii}
K absorptions of the objects when analysing individual spectra, being particularly noticeable for HD~85905. Equivalent widths of the circumstellar components and column densities were also computed for every date, averaging the spectra of each night (Table \ref{tab:alldata}), as no significant variations were observed at short timescales. The bottom half of Table \ref{tab:paramsfromews} shows values of the mean, median and standard deviation of the equivalent widths derived from the set of individual spectra for each star. It is clear that the dispersion of values around the mean is substantially larger, in relative terms, for HD~85905. Figure \ref{maxvars} shows the spectra that exhibit the greatest variations with respect to the average spectra for each star.

\begin{table}[t]
    \caption{Average parameters from the equivalent widths.}
    \label{tab:paramsfromews}
    \centering
    \setlength{\tabcolsep}{4.5 pt}
    \begin{tabular}{l c c c}
    \hline\hline
    \noalign{\smallskip}
                &  HD~36546          & HD~42111           & HD~85905     \\
    \noalign{\smallskip}
    \hline
    \noalign{\smallskip}
    \multicolumn{4}{l}{Average spectrum} \\
    \noalign{\smallskip}
    \hline
    \noalign{\smallskip}
    EW$_{\rm average}$ (m\AA)      & 30.0$\pm$0.3       & 339.0$\pm$1.0       & 70.7$\pm$7.0 \\ 
    N$_{\rm average}$  (cm$^{-2}$) & 3.21$\times10^{11}$ & 3.63$\times10^{12}$ & 7.57$\times10^{11}$\\
    \noalign{\smallskip}
    \hline
    \noalign{\smallskip}
    \multicolumn{4}{l}{Set of individual spectra} \\ 
    \noalign{\smallskip}
    \hline
    \noalign{\smallskip}
    mean(EW$_i$)    (m\AA)      & 30.0               & 329.5        & 73.0                    \\
    median(EW$_i$)  (m\AA)      & 30.3               & 327.0        & 76.5                    \\
    $\sigma$(EW$_i$) (m\AA)      & 1.0                &   28.2       &   15.6                  \\
    \noalign{\smallskip}
   \hline        
   \end{tabular}
\end{table}

\subsection{HD~85905}
\label{sec:hd85905}

The pronounced variability of HD~85905 motivated us to perform a dedicated analysis of this object. The observed variations are complex and appear to result from a combination of multiple components whose nature remains uncertain. In the following sections, we reassess the stellar parameters according to the binary configuration of this object, discuss the results obtained from the spectral fittings and the spectral energy distribution (SED), and describe the variability patterns in detail.

\subsubsection{Binarity} 
\label{sec:binarity}

\begin{figure}
        \centering
            \includegraphics[width=\linewidth]{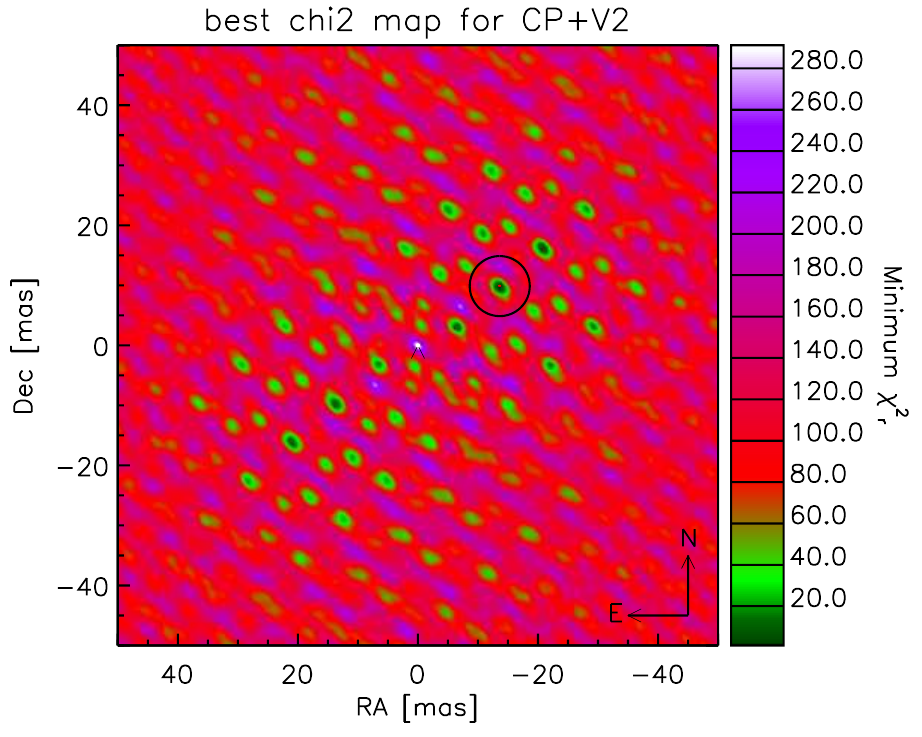}
        \caption{Normalized $\chi^2$ map of the combined $V^2$+CP analysis for the VLTI/PIONIER observation. The white star at the centre marks the (fixed) position of the primary star in the binary model, while the black circle indicates the position of the minimum in the map, i.e., the most likely position for the companion.}
        \label{fig:pionier}
    \end{figure}

To affirm the presence of a stellar companion based on the PIONIER data, we use the squared visibilities ($V^2$) and closure phases (CP) and in a combined way. As in \citet{Marion14}, we consider a binary star model with the primary at the centre of the search region and an off-axis companion of varying contrast at each point of the search region. In the present case, we can safely assume that both the primary and the secondary stars are unresolved, as our target star has an estimated angular diameter of about $0.22$~mas based on surface brightness relationships \citep{Kervella04}. We then compute the $V^2$ and CP for each binary model and derive a combined goodness of fit that we normalise and collapse along the contrast axis to keep only the best-fitting companion contrast (i.e., minimum $\chi^2$ value) at each position in the search region. The resulting $\chi^2$ map, displayed in Fig.~\ref{fig:pionier}, can then be used to derive the probability for the single-star model to adequately represent the data based on the $\chi^2$ distribution under a Gaussian noise assumption. With a normalised $\chi^2$ larger than 280, the single-star model can be rejected with a very high confidence. 

While the $\chi^2$ map shows a series of local minima, only one position leads to a normalised $\chi^2$ lower than 5, so that the best-fit position can be considered as unambiguous. The best-fit binary model is displayed as solid lines in Fig.~\ref{fig:pionier_data}, showing an excellent agreement with the data\footnote{In some cases, the binary model seems not to reproduce the measured CP, but this is due to the 180\degr wrapping in CP measurements, which also explains the very large error bars on some CP.}. The binary fit then yields the following results: angular separation of $16.8 \pm 0.3$ mas, position angle of $-54\fdg1 \pm 0\fdg5$, and $H$-band contrast of $0.855 \pm 0.076$. Based on the derived angular separation and on the distance to the system, and assuming a circular face-on orbit, the semi-major axis would be about 2.6 au, and the period around 2.2 years. From the results of the PIONIER observation alone, which is obtained in $H$-band, the companion could be slightly evolved, as the A1~IV primary, and then be of a similar spectral type, or, if it were on the main sequence, it could be a star of spectral type $\sim$B4~V; this second possibility is inconsistent when the observed luminosity of the object is taken into account.

As supplementary information, it is interesting to note that data from \textit{Gaia} DR2 and DR3 already provided some hints about the potential binarity. The {\tt RUWE} (re-normalised unit weight error) parameter increased from 0.8653 (DR2) to 1.2976 (DR3). RUWE informs about the astrometric deviation from using single-source solutions, the flag warning about potential ``bad'' solutions having been set to 1.4 (DR2) and 1.25 (EDR3) \citep{Castro24}. The values for the DR3 parameters {\tt{ipd\_gof\_harmonic\_phase}} (150), and {\tt{ipd\_frac\_multi\_peak}} (1), according to the \textit{Gaia} DR3 documentation\footnote{\url{https://gea.esac.esa.int/archive/documentation/GDR3/Gaia_archive/}, Chapter 20, section 20.1.1.}, suggests that the combination of the two values would be consistent with the hypothesis of a binary scenario.

\subsubsection{Stellar parameters}
\label{sec:parameters}

In view of the results presented in the previous section, the spectroscopic determination by \cite{Rebollido20}, done under the assumption that the star was a single object (see Table \ref{tab:parameters}), must obviously be revised. An interesting fact of both the spectroscopic and photometric analyses is that a single-temperature, single-gravity model provides a set of parameters that is apparently self-consistent, not showing any hint that a combination of two different stellar models is required to fit the spectrum and the SED. However, the luminosity, $L_*/L_\odot\!=\!112.6\pm10.0$, computed by the integration of the dereddened best-fitting model, using the distance from the \textit{Gaia} DR3 parallax, $\varpi\!=\!5.4555 \pm 0.0433$ mas, does not correspond to that of a single star of spectral type A1 IV located at that distance, but to a much brighter object\footnote{The error bar assigned to the luminosity is a conservative estimate accounting for the small uncertainty
introduced by the parallax and the normalization of the model to the 
observed SED.} (see Appendix \ref{app:photometryandhr} for details).

A solution, congruent with the possibility suggested by the PIONIER observation, namely that of an almost-identical component binary, would be to consider that the binary is made of two coeval objects of similar spectral types. The total luminosity, resulting of adding up the individual values, would match the observed one, with both stars being slightly off the main sequence with parameters consistent with the observed spectral type A1 IV. 

\begin{table}[t]
    \caption{Parameters for the HD~85905 binary solution}
    \label{tab:binary}
    \centering
    \setlength{\tabcolsep}{9.0 pt}
    \begin{tabular}{l c c}
    \hline\hline
    \noalign{\smallskip}
      & Component 1 & Component 2  \\ 
    \noalign{\smallskip}
    \hline
    \noalign{\smallskip}
 $T_{\rm eff}$ (K) &  9010 & 9105 \\
 $\log g$          &  3.82 & 3.89 \\
 $L_*$ ($L_\odot$) &  60.3 & 52.5 \\
 $M$ ($M_\odot$)   &  2.46 & 2.40 \\
 $H$ (mag)         &  0.479& 0.298 \\
 Age (Myr)         &  530  & 530   \\
   \noalign{\smallskip}
   \hline
   \end{tabular}
   \tablefoot{No uncertainties are given associated to any of the quantities, since they are extracted from exact values contained in the PARSEC 2.1s isochrones. The vertical error bars plotted in Fig. \ref{fig:sedandevolution} on the blue dots representing the components C1 and C2 have conservative values to account for the uncertainty in the luminosity.}
\end{table}   

We explored the parameter space for solutions with temperatures around and 
close to 9040 K --which was the temperature determined from the spectroscopic analysis-- using the data contained in the PARSEC 2.1s evolutionary tracks and isochrones\footnote{\url{https://stev.oapd.inaf.it/PARSEC/tracks_v12s.html}} 
\citep{Bressan12}. The best result is the sets of parameters given in Table \ref{tab:binary}, fulfilling the constraints: (i) both stars have temperatures close to that of the one-temperature fit, (ii) the sum of their luminosities matches the observed one, (iii) the contrast between the fluxes in $H$-band is 0.855, and (iv) the stars are coeval. Figure \ref{fig:sedandevolution} shows in the top panels
the HR diagrams luminosity-temperature and gravity-temperature with the 
putative the position of the star according to the one-temperature fit (red dot) and the locations proposed for the two-component solution, labelled C1 and C2. In the bottom panel, the two Kurucz synthetic models, reddened with A$_V\!=\!0.237$, and their sum, plotted in green (C1), magenta (C2) and grey (C1+C2), are shown together with the observed photometry, the HST/STIS spectrum and the IUE spectrum SWP~34794 extracted from the INES database\footnote{\url{https://sdc.cab.inta-csic.es/cgi-ines/IUEdbsMY}}.

\subsubsection{Variability}

\begin{figure*}[t]
    \centering
    \includegraphics[width=1\linewidth]{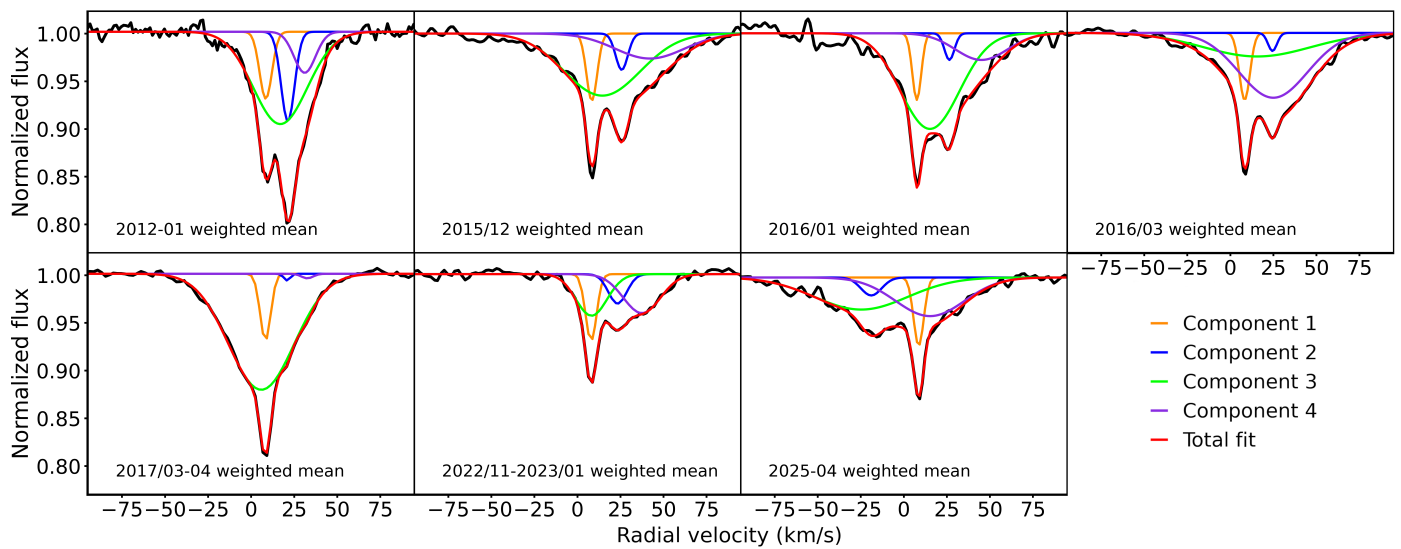}
    \caption{Gaussian fits to the circumstellar components of the Ca {\sc ii} K line
    for HD~85905. The observed profiles, plotted in black are the result of dividing the corresponding weighted mean spectra and the photospheric profile. Four Gaussian curves have been fitted to each profile, with the results of the fits --the addition of the four Gaussians-- plotted in red. The orange component does not vary significantly through the different epochs.}
    \label{fig:gaussfits}
\end{figure*}

\begin{table*}[t]
\caption{HD~85905: Parameters of the components (Fig. \ref{fig:gaussfits})}
    \label{tab:gaussfits}
    \setlength{\tabcolsep}{5.8pt}
    \begin{tabular}{c|ccc|crc|crc|ccr}
    \hline\hline
    \noalign{\smallskip}
    & \multicolumn{3}{c|}{Component 1} &   
      \multicolumn{3}{c|}{Component 2} &
      \multicolumn{3}{c|}{Component 3} &
      \multicolumn{3}{c}{Component 4} \\
   \noalign{\smallskip}    
    \hline
   \noalign{\smallskip}    
    Date & $A$ & $\mu$ & $\sigma$ & $A$ & \multicolumn{1}{c}{$\mu$} & $\sigma$ & $A$ & \multicolumn{1}{c}{$\mu$} & $\sigma$ & $A$ & $\mu$ & \multicolumn{1}{c}{$\sigma$} \\
         &     & (km/s) & (km/s) &   &(km/s) & (km/s) &  & (km/s) & (km/s)
         &     & (km/s) & (km/s) \\
    \noalign{\smallskip}   
    \hline
    \noalign{\smallskip}   
    2012/01 &  0.070 & 8.49 & 3.59 &  0.094 & 21.32 & 3.91 &  0.098 & 16.95 & 15.69 &  0.043 & 31.10 & 6.22 \\
    2015/12 &  0.070 & 8.43 & 3.11 &  0.038 & 25.83 & 3.53 &  0.065 & 14.52 & 22.00 &  0.026 & 42.12 & 20.00 \\
    2016/01 &  0.070 & 7.54 & 2.58 &  0.028 & 26.49 & 2.91 &  0.100 & 15.08 & 17.00 &  0.029 & 44.86 & 15.00 \\
    2016/03 &  0.070 & 8.31 & 3.09 &  0.019 & 24.49 & 2.62 &  0.025 & 15.56 & 31.11 &  0.068 & 24.97 & 19.86\\
    2017/03-04 & 0.070 & 8.29 & 3.07 &  0.007 & 20.86 & 1.89 &  0.122 & 5.91 & 19.06 & 0.005 & 32.61 & 3.35 \\
    2022/11-2023/01 & 0.070 & 7.91 & 3.20 &  0.031 & 23.31 & 5.36 &  0.044 & 8.15 & 9.02 &  0.041 & 37.43 & 10.78\\
    2025/04 & 0.070 & 8.63 & 2.95 &  0.019 & $-18.90$ & 6.00 &  0.033 & $-24.78$ & 27.00 & 0.040 & 15.24 & 20.00 \\
    \noalign{\smallskip}    
    \hline
    \noalign{\smallskip}
    \end{tabular}
    \tablefoot{$A$ is the depth in units of normalized fluxes measured from
    the continuum at intensity 1.0, $\mu$ and $\sigma$ are the position and standard deviation in km/s, respectively. The relationship between the full width at half maximum (FWHM) of the Gaussian and $\sigma$ is: FWHM=$2\sqrt{2\,\ln 2}\,\sigma$.
    }
\end{table*}

 The circumstellar profiles for most epochs could be fitted with a minimum of four Gaussian components. Figure \ref{fig:gaussfits} shows the mean circumstellar profiles, plotted in black, of seven epochs in the interval 2012--2025, with the different components separated by colours, and their combination plotted in red.  Table \ref{tab:gaussfits} gives the depth, $A$, position, $\mu$, and standard deviation $\sigma$ of the individual components. Throughout the observed spectra, dramatic variations are apparent, with a stable component present in all epochs, plotted in orange -labelled as "component 1"- remaining roughly constant in EW, depth, radial velocity and width. Other transient components appear and disappear both blue and red-shifted with respect to the radial velocity of the star, following a complex pattern. The depth of component 1 has been kept constant, allowing the remaining of that component to vary. The mean values for the position and standard deviation are $\bar{\mu_1} = 8.23 \pm 0.38$ km/s, $\bar{\sigma_1}\!= 3.08 \pm 0.30$ km/s, respectively.

Fig. \ref{HD85905all} shows in more detail the circumstellar components of the Ca {\sc ii} K line for HD~85905 in multiple epochs. The panel on the left shows spectra extracted from papers and scanned: 1987/05 \citep{Lagrange90}, 1997/01 and 1997/11 \citep{Welsh98}. Those from 2004/12 and 2005/02 are archival observations \citep{Redfield07}. These spectra are shown for comparison, and have not been used in the analysis carried out in this paper. The three panels on the right show the circumstellar profiles of observations obtained in the period 2012--2025. Red and blue labels in Fig. \ref{HD85905all} alternate to separate observations of different campaigns, and help to make comparisons of the differences between consecutive runs. In addition to the dates, the spectrographs used are specified in the labels. The depths, widths and global shapes of the circumstellar component vary without any apparent pattern. The stable component at $\sim\!8.23$ km/s is the only feature that remains roughly unchanged, whereas around 25 km/s there is a feature that appears and disappears in an unpredictable manner.

\begin{figure}[!ht]
    \centering
    \includegraphics[width=0.97\linewidth]{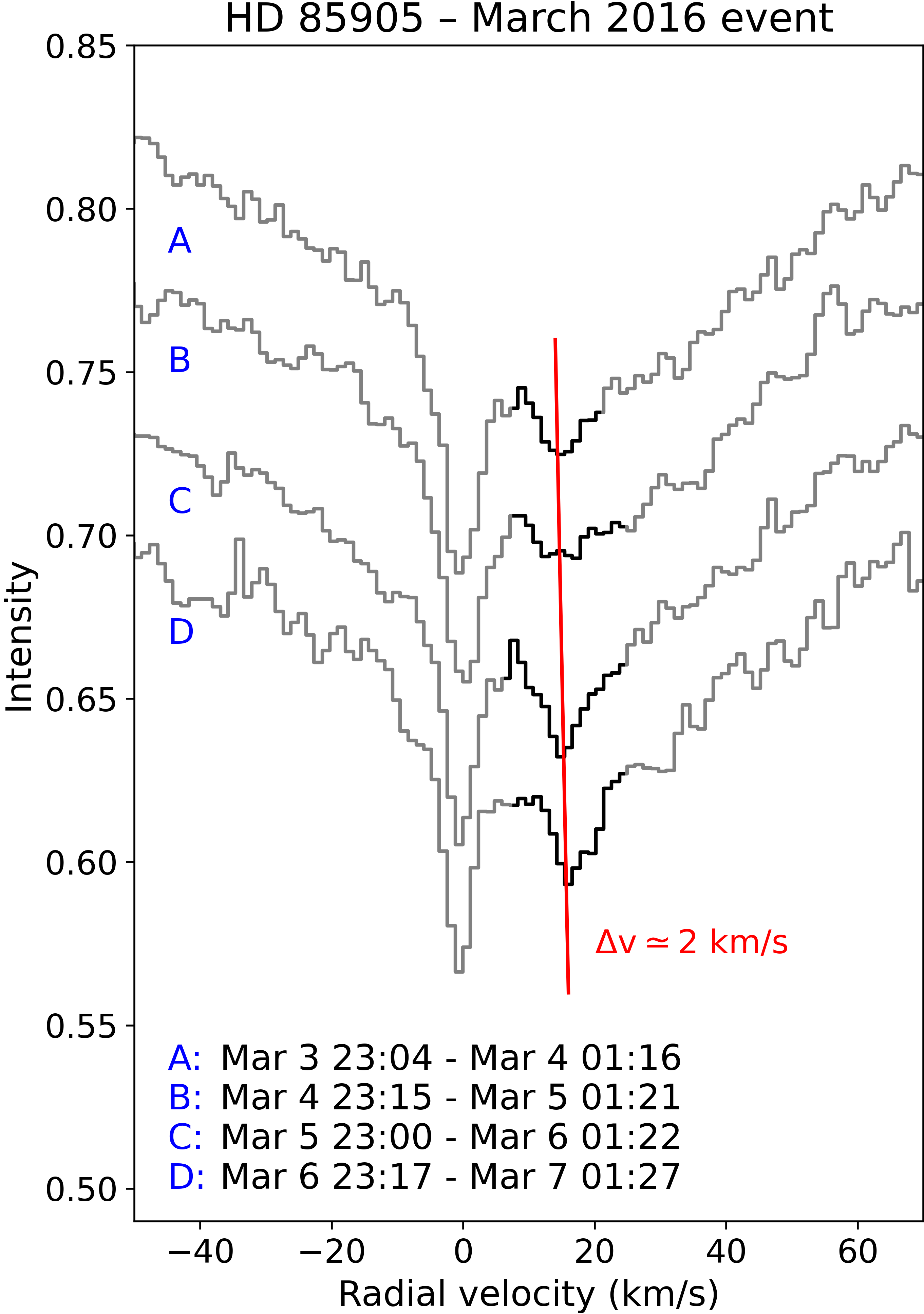}
    \caption{Example of variability in the circumstellar Ca {\sc ii} K component of HD~85905 on timescales of days. The plot shows the weighted mean of spectra obtained close in time at the beginning of four consecutive nights starting around UT 23:00 on 3–6 March 2016, with 4, 4, 3, and 3 spectra contributing to the respective means. Variability in the component at $\sim\!23$ km/s is clearly visible; its minima, joined by a red line, are redshifted
    by $\sim$2 km/s from spectrum A to D, hinting to a 
    mild acceleration of the clump of gas producing that absorption.}
    \label{fig:event2016}
\end{figure}

Figure \ref{fig:event2016} shows an example of the variability of HD~85905 on a daily timescale. The plot shows the mean of four sets of spectra obtained in intervals of $\sim\!2.5$ h during four consecutive nights. The stable component of the circumstellar absorption at $\sim\!8.23$ km/s appears unchanged, whereas a component at $\sim\!23$ km/s shows a clear variation. 
The red vertical line in the figure connects the maximum depth of the variable feature observed in each spectra, spanning from 14 to 16 km/s in the stellar reference frame, and suggesting a mild acceleration of the gas responsible for this component.

\section{Discussion}
\label{sec:discussion}

Detecting exocometary phenomena through spectroscopic monitoring is inherently challenging. In addition to requiring high-resolution, high signal-to-noise ratio spectra, it also demands a temporal cadence that is, with few exceptions, difficult to achieve. A good example of a successful monitoring is that for $\beta$ Pic \citep{Kennedy18} where subtle exocometary events --see Fig.~4 of that paper-- were detected and studied in depth; in particular the one on MJD=54913  required the analysis of 61 HARPS spectra obtained during an interval of just 2 hours. Such an optimal time-intensive strategy, feasible for that object given its brightness and the well-know high exocometary activity, is risky for other exocomet-host candidates, as the transient nature of these events implies a high likelihood of potential non-detections. 

While all three target stars in this paper (HD~36546, HD~42111 and HD~85905) have been classified in the past as exocomet host stars, the variations detected for HD~36546 and HD~42111 in the set of spectroscopic observations studied in this paper are very small. In the case of HD~36546, the shape of the absorption had already been reported as asymmetric \citep{Rebollido20}, with previous variations observed red-shifted with respect to the radial velocity of the star, consistent with an exocometary origin. The variations observed here are mostly related to the depth of the circumstellar component, and only within a
few percents of the EW of the average spectra.
HD~42111 is a known shell star \citep{Grady96} that had previously shown low velocity variations in the circumstellar absorption \citep{Welsh13}. However, the variations observed in this work are, as in the case of HD~36546, very small (see Table \ref{tab:paramsfromews}). There is no further evidence of exocomet-like variability for either of these stars.

Regarding HD~85905, both from this work and previous results in the literature, the extreme variability spotted in the circumstellar Ca {\sc ii} K component in time scales of days, weeks and months, qualitatively consistent in some aspects with the FEB scenario, cannot be entirely attributed, beyond any doubt, to exocometary activity, considering the discovery that the star is actually
a binary. This scenario would be consistent with reported binarity fractions for A-type stars. According to \cite{duchene13}, the multiplicity fraction of main-sequence early type stars should be above 50-60\%, reaching almost 80\% for B-type stars.

\begin{figure}[t]
    \centering
    \includegraphics[width=1\linewidth]{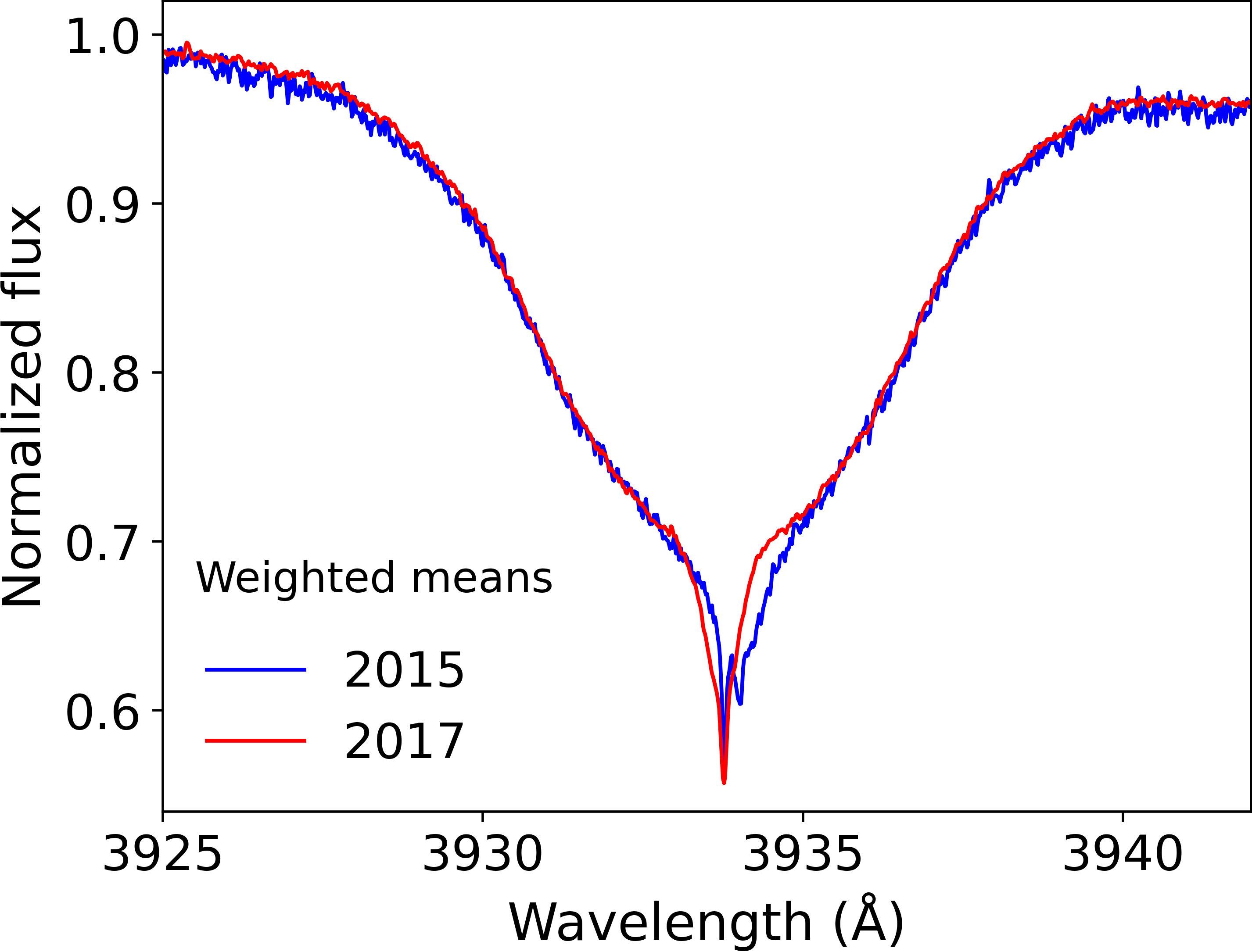}
    \caption{Full photospheric profiles of the Ca {\sc ii} K line
    -with the corresponding circumstellar absorptions at the bottom- extracted
    from the weighted mean spectra of HD~85905 corresponding to the
    2015 and 2017 campaigns. No differences are observed in the width
    and shape of the lines.}
    \label{fig:phot2015-2017}
\end{figure}

To ascertain the role of binarity in the variable circumstellar
features of HD~85905, there are many open questions regarding its architecture that can only be answered after an intensive spectroscopic monitoring of the star and further interferometric observations to characterize the orbit.
Figure \ref{fig:phot2015-2017} shows the photospheric profiles --with the
corresponding circumstellar components superimposed on the bottom-- of the Ca {\sc ii} K line from the weighted mean spectra for 2015 and 2017. The width and shape of the photospheric line in both epochs are virtually the same, 
whereas if the minima observed in the 2015 spectrum are interpreted as arising from the two different components of the binary system, one from each star, then appreciable changes in the profile width would be expected between the 2015 and 2017 spectra, which is not the case. This behaviour is shared by all photospheric lines. The observation that the stable circumstellar component does not change its position and it is always centred at the same radial velocity could indicate that it originates in a circumbinary disk.

The fact that the shape and width of the photospheric Ca {\sc ii} K line seem to remain stable could be explained claiming that the orbit is actually face-on. However, this seems to be quite unlikely: the width of the photospheric Ca {\sc ii} K profile at half intensity is around 460 km/s, which implied a value of $\varv \sin i\!\simeq\!285$ km/s to reproduce not only the width of this line, but also the profiles of other photospheric features when a single-$T$, single-$\log g$ fit was carried out to fit the observed spectrum \citep{Rebollido20}. If the inclination of the orbit were small -and assuming that that rotation axes of the stars are perpendicular to the plane of the orbit- such a value of $\varv \sin i$ would be totally unrealistic. Discarding the hypothesis of a low inclination for the orbit, a substantial difference between the projected rotational speeds, with the spectrum dominated by the fast rotator, could help explain this behaviour of the spectrum.

Additional hypotheses should be assessed in order to confirm or discard the potential exocometary origin of the variability observed in HD~85905. In addition to the results of the photometric analysis, the variability in the position and depth of the circumstellar components, and in particular the appearance and disappearance of the component redwards of the stable component could be compatible with its origin being located closer to the companion, in a configuration qualitatively similar to that in HR 10 \citep{Montesinos19}; this, of course, does not rule out the possibility that the origin be linked to the primary component of the binary. Despite the fairly large number of spectra collected, the irregular separations between campaigns does not allow to extract any meaningful information to build a radial velocity curve. 

Another possibility can include pulsation modes of the star altering its circumstellar environment. This has been observed in the case of shell stars \citep{Eiroa21}. However, that explanation was discarded as there is no evidence of variability in other photospheric features, such as the Balmer lines. The significant variability observed also rules out the possibility of an ISM origin for the non-photospheric absorptions. 

Since all variable features observed are not explained by circumstellar gas around both components of the binary, the potential presence of exocomets around HD~85905 cannot be discarded. Figure \ref{fig:event2016} shows four mean spectra of the Ca {\sc ii} K line obtained during approximately 2 hours at the beginning of four consecutive nights on 3-6 March 2016; the slight shift to the red, $\Delta v\!\simeq\!2$ km/s in 4 days, of the component at $\sim\!23$ km/s could suggest a mild acceleration of the absorbing gas, that can be compared with the results obtained by \cite{Kennedy18} for $\beta$ Pic,
although it could also be related with the orbital motion of the binary. However, as we pointed out above, testing the exocomet hypothesis is very challenging due to the sporadic nature of these events, particularly before the orbit of the binary is properly parametrized, and therefore more data will be necessary. For instance, there are no reported detections of variable gas in UV wavelengths, or sub-mm observations that could point to high velocity exocomets \citep{Kiefer14b, Vrignaud2025}. Photometric time series could also help confirm this scenario, the data obtained by TESS, although publicly available, remain unpublished, and its analysis is out of the scope of this work. 

\section{Conclusions}
\label{sec:conclusions}
We present high-resolution observations of two main-sequence and a subgiant star aiming to detect exocometary-like features. All three objects in this study have shown in the past circumstellar features with different degrees of variability, which imply that changes in the material surrounding the objects have occurred. Two objects, HD~36546 and HD~42111 do not show significant variations in the sets of observations studied in this paper whereas 
HD~85905 presents signatures that resemble those of the $\beta$ Pic absorptions, as in \cite{Ferlet87}. Fig. \ref{HD85905all} shows that the variations seem to take place more frequently at a redshifted position relative to the star radial velocity, although some sporadic --but intense-- episodes, have also been detected at blue wavelengths. The explanation of the variability of this star is a question that remains open.
The fact that HD~85905 is a binary makes the problem even more complex, because the question to be answered is whether the variable circumstellar absorptions observed in this object, and the peculiar behaviour of its spectra, can be attributed to the binary orbit alone, or whether they have an exocometary origin.

The hypothesis of the existence of cometary material around the three stars, and in particular around HD~85905 is not yet firm, because there is the possibility that the circumstellar absorptions might be due to dust grain evaporation and/or grain-grain collisions in a disk or envelope. Intensive well-sampled systematic observations on different timescales are required to confirm or dismiss the presence of exocomets close to the stars.

\begin{acknowledgements}
We thank the referee for the constructive report, and the suggestions and comments, which have undoubtedly contributed to improving the original manuscript. We are grateful to Prof. Seth Redfield (Wesleyan University, USA) for providing the spectra of HD~85905 obtained with the McD2.7 telescope in 2004 and 2005, and to Dr Jes\'us Ma\'{\i}z-Apell\'aniz (CAB, CSIC-INTA) for observing the star with Mercator/HERMES in April 2025. We are also grateful to Dr H\'ector C\'anovas for helping us with the interpretation of some \textit{Gaia} parameters for HD~85905. 
We would like to thank Sergio Suárez and Antonio Parras for their ongoing assistance, availability, kindness, and impeccable maintenance of the CAB's computer system. B. M. and I. R. acknowledge the funding by grant PID2021-127289-NB-I00 from MCIN/AEI/10.13039/501100011033/ and FEDER. P. M.-C. acknowledges the funding by grant PID2022-137980NB-I00. P. C. acknowledges financial support from the Spanish Ministry of Science and Innovation/State Agency of Research MCIN/AEI/10.13039/501100011033 through the Spanish Virtual Observatory project PID2023-146210NB-I00.
S. E. is supported by NASA through grants 80NSSC21K0394, 80NSSC23K1473, and 80NSSC23K0288. This publication makes use of VOSA, developed under the Spanish Virtual Observatory (https://svo.cab.inta-csic.es) project funded by MCIN/AEI/10.13039/501100011033/ through grant PID2020-112949GB-I00. VOSA has been partially updated by using funding from the European Union's Horizon 2020 Research and Innovation Programme, under Grant Agreement no. 776403 (EXOPLANETS-A). This research has made use of the SIMBAD database, operated at CDS, Strasbourg, France; and also 
of Starlink software \citep{Currie14}, currently supported by the East Asian Observatory.
\end{acknowledgements}

\bibliographystyle{aa}
\bibliography{bibliography.bib}

@ARTICLE{Rebollido20,
       author = {{Rebollido}, I. and {Eiroa}, C. and {Montesinos}, B. and {Maldonado}, J. and {Villaver}, E. and {Absil}, O. and {Bayo}, A. and {Canovas}, H. and {Carmona}, A. and {Chen}, Ch. and {Ertel}, S. and {Henning}, Th. and {Iglesias}, D.~P. and {Launhardt}, R. and {Liseau}, R. and {Meeus}, G. and {Mo{\'o}r}, A. and {Mora}, A. and {Olofsson}, J. and {Rauw}, G. and {Riviere-Marichalar}, P.},
        title = "{Exocomets: A spectroscopic survey}",
      journal = {\aap},
         year = 2020,
        month = jul,
       volume = {639},
          eid = {A11},
        pages = {A11},
          doi = {10.1051/0004-6361/201936071},
archivePrefix = {arXiv},
       eprint = {2003.11084},
 primaryClass = {astro-ph.SR},
       adsurl = {https://ui.adsabs.harvard.edu/abs/2020A&A...639A..11R}
}

@article{Kiefer14b,
  TITLE = {{Two families of exocomets in the $\beta$ Pictoris system}},
  AUTHOR = {Kiefer, F. and Lecavelier Des Etangs, A. and Boissier, J. and Vidal-Madjar, A. and Beust, H. and Lagrange, A. -M. and H{\'e}brard, G. and Ferlet, R.},
  URL = {https://insu.hal.science/insu-03645266},
  JOURNAL = {{Nature}},
  PUBLISHER = {{Nature Publishing Group}},
  VOLUME = {514},
  PAGES = {462-464},
  YEAR = {2014b},
  DOI = {10.1038/nature13849},
  HAL_ID = {insu-03645266},
  HAL_VERSION = {v1},
}

@ARTICLE{Worthen24a,
       author = {{Worthen}, Kadin and {Chen}, Christine H. and {Brittain}, Sean D. and {Lu}, Cicero X. and {Rebollido}, Isabel and {Brennan}, Aoife and {Matr{\`a}}, Luca and {Melis}, Carl and {Delgado}, Timoteo and {Roberge}, Aki and {Mazoyer}, Johan},
        title = "{Vertical Structure of Gas and Dust in Four Debris Disks}",
      journal = {\apj},
         year = 2024,
        month = feb,
       volume = {962},
       number = {2},
          eid = {166},
        pages = {166},
          doi = {10.3847/1538-4357/ad1511},
archivePrefix = {arXiv},
       eprint = {2312.09106},
 primaryClass = {astro-ph.EP},
       adsurl = {https://ui.adsabs.harvard.edu/abs/2024ApJ...962..166W}
}

@ARTICLE{Beust90,
       author = {{Beust}, H. and {Lagrange-Henri}, A.~M. and {Vidal-Madjar}, A. and {Ferlet}, R.},
        title = "{The beta Pictoris circumstellar disk. X. Numerical simulations of infalling evaporating bodies.}",
      journal = {\aap},
         year = 1990,
        month = sep,
       volume = {236},
        pages = {202},
       adsurl = {https://ui.adsabs.harvard.edu/abs/1990A&A...236..202B}
}

@ARTICLE{Somerville88,
       author = {{Somerville}, W.~B.},
        title = "{The doublet-ratio method for interstellar abundances}",
      journal = {The Observatory},
         year = 1988,
        month = apr,
       volume = {108},
        pages = {44-49},
       adsurl = {https://ui.adsabs.harvard.edu/abs/1988Obs...108...44S}
}

@ARTICLE{Ferlet87,
       author = {{Ferlet}, R. and {Hobbs}, L.~M. and {Vidal-Madjar}, A.},
        title = "{The beta Pictoris circumstellar disk. V. Time variations of the Ca II-K line.}",
      journal = {\aap},
         year = 1987,
        month = oct,
       volume = {185},
        pages = {267-270},
       adsurl = {https://ui.adsabs.harvard.edu/abs/1987A&A...185..267F}
}

@ARTICLE{Zieba19,
       author = {{Zieba}, S. and {Zwintz}, K. and {Kenworthy}, M.~A. and {Kennedy}, G.~M.},
        title = "{Transiting exocomets detected in broadband light by TESS in the {\ensuremath{\beta}} Pictoris system}",
      journal = {\aap},
         year = 2019,
        month = may,
       volume = {625},
          eid = {L13},
        pages = {L13},
          doi = {10.1051/0004-6361/201935552},
archivePrefix = {arXiv},
       eprint = {1903.11071},
 primaryClass = {astro-ph.SR},
       adsurl = {https://ui.adsabs.harvard.edu/abs/2019A&A...625L..13Z}
}

@ARTICLE{Strom20,
       author = {{Str{\o}m}, Paul A. and {Bodewits}, Dennis and {Knight}, Matthew M. and {Kiefer}, Flavien and {Jones}, Geraint H. and {Kral}, Quentin and {Matr{\`a}}, Luca and {Bodman}, Eva and {Capria}, Maria Teresa and {Cleeves}, Ilsedore and {Fitzsimmons}, Alan and {Haghighipour}, Nader and {Harrison}, John H.~D. and {Iglesias}, Daniela and {Kama}, Mihkel and {Linnartz}, Harold and {Majumdar}, Liton and {de Mooij}, Ernst J.~W. and {Milam}, Stefanie N. and {Opitom}, Cyrielle and {Rebollido}, Isabel and {Rogers}, Laura K. and {Snodgrass}, Colin and {Sousa-Silva}, Clara and {Xu}, Siyi and {Lin}, Zhong-Yi and {Zieba}, Sebastian},
        title = "{Exocomets from a Solar System Perspective}",
      journal = {\pasp},
         year = 2020,
        month = oct,
       volume = {132},
       number = {1016},
          eid = {101001},
        pages = {101001},
          doi = {10.1088/1538-3873/aba6a0},
archivePrefix = {arXiv},
       eprint = {2007.09155},
 primaryClass = {astro-ph.EP},
       adsurl = {https://ui.adsabs.harvard.edu/abs/2020PASP..132j1001S}
}

@ARTICLE{Montesinos19,
       author = {{Montesinos}, B. and {Eiroa}, C. and {Lillo-Box}, J. and {Rebollido}, I. and {Djupvik}, A.~A. and {Absil}, O. and {Ertel}, S. and {Marion}, L. and {Kajava}, J.~J.~E. and {Redfield}, S. and {Isaacson}, H. and {C{\'a}novas}, H. and {Meeus}, G. and {Mendigut{\'\i}a}, I. and {Mora}, A. and {Rivi{\`e}re-Marichalar}, P. and {Villaver}, E. and {Maldonado}, J. and {Henning}, T.},
        title = "{HR 10: a main-sequence binary with circumstellar envelopes around both components. Discovery and analysis}",
      journal = {\aap},
         year = 2019,
        month = sep,
       volume = {629},
          eid = {A19},
        pages = {A19},
          doi = {10.1051/0004-6361/201936180},
archivePrefix = {arXiv},
       eprint = {1907.12441},
 primaryClass = {astro-ph.SR},
       adsurl = {https://ui.adsabs.harvard.edu/abs/2019A&A...629A..19M}
}

@ARTICLE{Eiroa21,
       author = {{Eiroa}, C. and {Montesinos}, B. and {Rebollido}, I. and {Henning}, Th. and {Launhardt}, R. and {Maldonado}, J. and {Meeus}, G. and {Mora}, A. and {Rivi{\`e}re-Marichalar}, P. and {Villaver}, E.},
        title = "{The A-shell star {\ensuremath{\phi}} Leo revisited: its photospheric and circumstellar spectra}",
      journal = {\aap},
         year = 2021,
        month = sep,
       volume = {653},
          eid = {A115},
        pages = {A115},
          doi = {10.1051/0004-6361/202141140},
archivePrefix = {arXiv},
       eprint = {2106.16229},
 primaryClass = {astro-ph.SR},
       adsurl = {https://ui.adsabs.harvard.edu/abs/2021A&A...653A.115E}
}

@ARTICLE{Lecavelier22,
       author = {{Lecavelier des Etangs}, Alain and {Cros}, Lucie and {H{\'e}brard}, Guillaume and {Martioli}, Eder and {Duquesnoy}, Marc and {Kenworthy}, Matthew A. and {Kiefer}, Flavien and {Lacour}, Sylvestre and {Lagrange}, Anne-Marie and {Meunier}, Nad{\`e}ge and {Vidal-Madjar}, Alfred},
        title = "{Exocomets size distribution in the {\ensuremath{\beta}} Pictoris planetary system}",
      journal = {Scientific Reports},
         year = 2022,
        month = apr,
       volume = {12},
          eid = {5855},
        pages = {5855},
          doi = {10.1038/s41598-022-09021-2},
archivePrefix = {arXiv},
       eprint = {2204.13618},
 primaryClass = {astro-ph.EP},
       adsurl = {https://ui.adsabs.harvard.edu/abs/2022NatSR..12.5855L}
}

@ARTICLE{Kiefer23,
       author = {{Kiefer}, F. and {Van Grootel}, V. and {Lecavelier des Etangs}, A. and {Szab{\'o}}, Gy. M. and {Brandeker}, A. and {Broeg}, C. and {Collier Cameron}, A. and {Deline}, A. and {Olofsson}, G. and {Wilson}, T.~G. and {Sousa}, S.~G. and {Gandolfi}, D. and {H{\'e}brard}, G. and {Alibert}, Y. and {Alonso}, R. and {Anglada}, G. and {B{\'a}rczy}, T. and {Barrado}, D. and {Barros}, S.~C.~C. and {Baumjohann}, W. and {Beck}, M. and {Beck}, T. and {Benz}, W. and {Billot}, N. and {Bonfils}, X. and {Cabrera}, J. and {Charnoz}, S. and {Csizmadia}, Sz. and {Davies}, M.~B. and {Deleuil}, M. and {Delrez}, L. and {Demangeon}, O.~D.~S. and {Demory}, B. -O. and {Ehrenreich}, D. and {Erikson}, A. and {Fortier}, A. and {Fossati}, L. and {Fridlund}, M. and {Gillon}, M. and {G{\"u}del}, M. and {Heng}, K. and {Hoyer}, S. and {Isaak}, K.~G. and {Kiss}, L.~L. and {Laskar}, J. and {Lendl}, M. and {Lovis}, C. and {Magrin}, D. and {Maxted}, P.~F.~L. and {Munari}, M. and {Nascimbeni}, V. and {Ottensamer}, R. and {Pagano}, I. and {Pall{\'e}}, E. and {Peter}, G. and {Piazza}, D. and {Piotto}, G. and {Pollacco}, D. and {Queloz}, D. and {Ragazzoni}, R. and {Rando}, N. and {Ratti}, F. and {Rauer}, H. and {Reimers}, C. and {Ribas}, I. and {Santos}, N.~C. and {Scandariato}, G. and {S{\'e}gransan}, D. and {Simon}, A.~E. and {Smith}, A.~M.~S. and {Steller}, M. and {Thomas}, N. and {Udry}, S. and {Walter}, I. and {Walton}, N.~A.},
        title = "{Hint of an exocomet transit in the CHEOPS light curve of HD 172555}",
      journal = {\aap},
         year = 2023,
        month = mar,
       volume = {671},
          eid = {A25},
        pages = {A25},
          doi = {10.1051/0004-6361/202245104},
archivePrefix = {arXiv},
       eprint = {2301.07418},
 primaryClass = {astro-ph.EP},
       adsurl = {https://ui.adsabs.harvard.edu/abs/2023A&A...671A..25K}
}

@ARTICLE{Lis19,
       author = {{Lis}, Dariusz C. and {Bockel{\'e}e-Morvan}, Dominique and {G{\"u}sten}, Rolf and {Biver}, Nicolas and {Stutzki}, J{\"u}rgen and {Delorme}, Yan and {Dur{\'a}n}, Carlos and {Wiesemeyer}, Helmut and {Okada}, Yoko},
        title = "{Terrestrial deuterium-to-hydrogen ratio in water in hyperactive comets}",
      journal = {\aap},
         year = 2019,
        month = may,
       volume = {625},
          eid = {L5},
        pages = {L5},
          doi = {10.1051/0004-6361/201935554},
archivePrefix = {arXiv},
       eprint = {1904.09175},
 primaryClass = {astro-ph.EP},
       adsurl = {https://ui.adsabs.harvard.edu/abs/2019A&A...625L...5L}
}

@ARTICLE{Val-Borro14,
       author = {{de Val-Borro}, M. and {Bockel{\'e}e-Morvan}, D. and {Jehin}, E. and {Hartogh}, P. and {Opitom}, C. and {Szutowicz}, S. and {Biver}, N. and {Crovisier}, J. and {Lis}, D.~C. and {Rezac}, L. and {de Graauw}, Th. and {Hutsem{\'e}kers}, D. and {Jarchow}, C. and {Kidger}, M. and {K{\"u}ppers}, M. and {Lara}, L.~M. and {Manfroid}, J. and {Rengel}, M. and {Swinyard}, B.~M. and {Teyssier}, D. and {Vandenbussche}, B. and {Waelkens}, C.},
        title = "{Herschel observations of gas and dust in comet C/2006 W3 (Christensen) at 5 AU from the Sun}",
      journal = {\aap},
         year = 2014,
        month = apr,
       volume = {564},
          eid = {A124},
        pages = {A124},
          doi = {10.1051/0004-6361/201423427},
archivePrefix = {arXiv},
       eprint = {1404.4869},
 primaryClass = {astro-ph.EP},
       adsurl = {https://ui.adsabs.harvard.edu/abs/2014A&A...564A.124D}
}

@ARTICLE{Rebollido22,
       author = {{Rebollido}, Isabel and {Ribas}, {\'A}lvaro and {de Gregorio-Monsalvo}, Itziar and {Villaver}, Eva and {Montesinos}, Benjam{\'\i}n and {Chen}, Christine and {Canovas}, H{\'e}ctor and {Henning}, Thomas and {Mo{\'o}r}, Attila and {Perrin}, Marshall and {Rivi{\`e}re-Marichalar}, Pablo and {Eiroa}, Carlos},
        title = "{The search for gas in debris discs: ALMA detection of CO gas in HD 36546}",
      journal = {\mnras},
         year = 2022,
        month = jan,
       volume = {509},
       number = {1},
        pages = {693-700},
          doi = {10.1093/mnras/stab2906},
archivePrefix = {arXiv},
       eprint = {2110.02308},
 primaryClass = {astro-ph.EP},
       adsurl = {https://ui.adsabs.harvard.edu/abs/2022MNRAS.509..693R}
}

@ARTICLE{Hobbs85,
       author = {{Hobbs}, L.~M. and {Vidal-Madjar}, A. and {Ferlet}, R. and {Albert}, C.~E. and {Gry}, C.},
        title = "{The gaseous component of the disk around beta Pictoris.}",
      journal = {\apjl},
         year = 1985,
        month = jun,
       volume = {293},
        pages = {L29-L33},
          doi = {10.1086/184485},
       adsurl = {https://ui.adsabs.harvard.edu/abs/1985ApJ...293L..29H}
}

@ARTICLE{Currie17,
       author = {{Currie}, Thayne and {Guyon}, Olivier and {Tamura}, Motohide and {Kudo}, Tomoyuki and {Jovanovic}, Nemanja and {Lozi}, Julien and {Schlieder}, Joshua E. and {Brandt}, Timothy D. and {Kuhn}, Jonas and {Serabyn}, Eugene and {Janson}, Markus and {Carson}, Joseph and {Groff}, Tyler and {Kasdin}, N. Jeremy and {McElwain}, Michael W. and {Singh}, Garima and {Uyama}, Taichi and {Kuzuhara}, Masayuki and {Akiyama}, Eiji and {Grady}, Carol and {Hayashi}, Saeko and {Knapp}, Gillian and {Kwon}, Jung-mi and {Oh}, Daehyeon and {Wisniewski}, John and {Sitko}, Michael and {Yang}, Yi},
        title = "{Subaru/SCExAO First-light Direct Imaging of a Young Debris Disk around HD 36546}",
      journal = {\apjl},
         year = 2017,
        month = feb,
       volume = {836},
       number = {1},
          eid = {L15},
        pages = {L15},
          doi = {10.3847/2041-8213/836/1/L15},
archivePrefix = {arXiv},
       eprint = {1701.02314},
 primaryClass = {astro-ph.EP},
       adsurl = {https://ui.adsabs.harvard.edu/abs/2017ApJ...836L..15C}
}

@ARTICLE{Welsh13,
       author = {{Welsh}, Barry Y. and {Montgomery}, Sharon},
        title = "{Circumstellar Gas-Disk Variability Around A-Type Stars: The Detection of Exocomets?}",
      journal = {\pasp},
         year = 2013,
        month = jul,
       volume = {125},
       number = {929},
        pages = {759},
          doi = {10.1086/671757},
       adsurl = {https://ui.adsabs.harvard.edu/abs/2013PASP..125..759W}
}

@ARTICLE{Cao24,
       author = {{Cao}, David and {Plavchan}, Peter and {Summers}, Michael},
        title = "{The Implications of 'Oumuamua on Panspermia}",
      journal = {\apj},
         year = 2024,
        month = aug,
       volume = {971},
       number = {2},
          eid = {160},
        pages = {160},
          doi = {10.3847/1538-4357/ad57b8},
archivePrefix = {arXiv},
       eprint = {2401.02390},
 primaryClass = {astro-ph.EP},
       adsurl = {https://ui.adsabs.harvard.edu/abs/2024ApJ...971..160C}
}

@ARTICLE{Lawson21,
       author = {{Lawson}, Kellen and {Currie}, Thayne and {Wisniewski}, John P. and {Tamura}, Motohide and {Augereau}, Jean-Charles and {Brandt}, Timothy D. and {Guyon}, Olivier and {Kasdin}, N. Jeremy and {Groff}, Tyler D. and {Lozi}, Julien and {Deo}, Vincent and {Vievard}, Sebastien and {Chilcote}, Jeffrey and {Jovanovic}, Nemanja and {Martinache}, Frantz and {Skaf}, Nour and {Henning}, Thomas and {Knapp}, Gillian and {Kwon}, Jungmi and {McElwain}, Michael W. and {Pyo}, Tae-Soo and {Sitko}, Michael L. and {Uyama}, Taichi and {Wagner}, Kevin},
        title = "{Multiband Imaging of the HD 36546 Debris Disk: A Refined View from SCExAO/CHARIS}",
      journal = {\aj},
         year = 2021,
        month = dec,
       volume = {162},
       number = {6},
          eid = {293},
        pages = {293},
          doi = {10.3847/1538-3881/ac2823},
archivePrefix = {arXiv},
       eprint = {2109.08984},
 primaryClass = {astro-ph.EP},
       adsurl = {https://ui.adsabs.harvard.edu/abs/2021AJ....162..293L}
}

@ARTICLE{Lisse17,
       author = {{Lisse}, C.~M. and {Sitko}, M.~L. and {Russell}, R.~W. and {Marengo}, M. and {Currie}, T. and {Melis}, C. and {Mittal}, T. and {Song}, I.},
        title = "{Spectral Evidence for an Inner Carbon-rich Circumstellar Belt in the Young HD 36546 A-star System}",
      journal = {\apjl},
         year = 2017,
        month = may,
       volume = {840},
       number = {2},
          eid = {L20},
        pages = {L20},
          doi = {10.3847/2041-8213/aa6ea3},
archivePrefix = {arXiv},
       eprint = {1704.06348},
 primaryClass = {astro-ph.SR},
       adsurl = {https://ui.adsabs.harvard.edu/abs/2017ApJ...840L..20L}
}

@ARTICLE{Welsh98,
       author = {{Welsh}, B.~Y. and {Craig}, N. and {Crawford}, I.~A. and {Price}, R.~J.},
        title = "{Beta Pic-like circumstellar disk gas surrounding HR 10 and HD 85905}",
      journal = {\aap},
         year = 1998,
        month = oct,
       volume = {338},
        pages = {674-682},
       adsurl = {https://ui.adsabs.harvard.edu/abs/1998A&A...338..674W}
}

@ARTICLE{Redfield07,
       author = {{Redfield}, Seth and {Kessler-Silacci}, Jacqueline E. and {Cieza}, Lucas A.},
        title = "{Spitzer Limits on Dust Emission and Optical Gas Absorption Variability around Nearby Stars with Edge-on Circumstellar Disk Signatures}",
      journal = {\apj},
         year = 2007,
        month = jun,
       volume = {661},
       number = {2},
        pages = {944-971},
          doi = {10.1086/517516},
archivePrefix = {arXiv},
       eprint = {astro-ph/0703089},
 primaryClass = {astro-ph},
       adsurl = {https://ui.adsabs.harvard.edu/abs/2007ApJ...661..944R}
}

@INPROCEEDINGS{Grady96,
       author = {{Grady}, C.~A. and {Perez}, M.~R. and {Talavera}, A.},
        title = "{The beta Pictoris Phenomenon in A-Shell Stars}",
    booktitle = {American Astronomical Society Meeting Abstracts},
         year = 1996,
       series = {American Astronomical Society Meeting Abstracts},
       volume = {185},
        month = dec,
          eid = {48.07},
        pages = {48.07},
       adsurl = {https://ui.adsabs.harvard.edu/abs/1994AAS...185.4807G}
}

@INCOLLECTION{Dones04,
       author = {{Dones}, L. and {Weissman}, P.~R. and {Levison}, H.~F. and {Duncan}, M.~J.},
        title = "{Oort cloud formation and dynamics}",
    publisher = {The University of Arizona Press, Tucson, Arizona},
    booktitle = {Comets II},
         year = 2004,
       editor = {{Festou}, Michel C. and {Keller}, H. Uwe and {Weaver}, Harold A.},
        pages = {153},
       adsurl = {https://ui.adsabs.harvard.edu/abs/2004come.book..153D}
}

@ARTICLE{Lagrange90,
       author = {{Lagrange-Henri}, A.~M. and {Ferlet}, R. and {Vidal-Madjar}, A. and {Beust}, H. and {Gry}, C. and {Lallement}, R.},
        title = "{Search for beta Pictoris-like star.}",
      journal = {\aaps},
         year = 1990,
        month = nov,
       volume = {85},
        pages = {1089},
       adsurl = {https://ui.adsabs.harvard.edu/abs/1990A&AS...85.1089L}
}

@ARTICLE{Kennedy18,
       author = {{Kennedy}, Grant M.},
        title = "{Exocomet orbit fitting: accelerating coma absorption during transits of {\ensuremath{\beta}} Pictoris}",
      journal = {\mnras},
         year = 2018,
        month = sep,
       volume = {479},
       number = {2},
        pages = {1997-2006},
          doi = {10.1093/mnras/sty1477},
archivePrefix = {arXiv},
       eprint = {1806.01284},
 primaryClass = {astro-ph.EP},
       adsurl = {https://ui.adsabs.harvard.edu/abs/2018MNRAS.479.1997K}
}

@ARTICLE{Castro24,
       author = {{Castro-Ginard}, Alfred and {Penoyre}, Zephyr and {Casey}, Andrew R. and {Brown}, Anthony G.~A. and {Belokurov}, Vasily and {Cantat-Gaudin}, Tristan and {Drimmel}, Ronald and {Fouesneau}, Morgan and {Khanna}, Shourya and {Kurbatov}, Evgeny P. and {Price-Whelan}, Adrian M. and {Rix}, Hans-Walter and {Smart}, Richard L.},
        title = "{Gaia DR3 detectability of unresolved binary systems}",
      journal = {\aap},
         year = 2024,
        month = aug,
       volume = {688},
          eid = {A1},
        pages = {A1},
          doi = {10.1051/0004-6361/202450172},
archivePrefix = {arXiv},
       eprint = {2404.14127},
 primaryClass = {astro-ph.GA},
       adsurl = {https://ui.adsabs.harvard.edu/abs/2024A&A...688A...1C}
}

@ARTICLE{Bayo08,
       author = {{Bayo}, A. and {Rodrigo}, C. and {Barrado Y Navascu{\'e}s}, D. and {Solano}, E. and {Guti{\'e}rrez}, R. and {Morales-Calder{\'o}n}, M. and {Allard}, F.},
        title = "{VOSA: virtual observatory SED analyzer. An application to the Collinder 69 open cluster}",
      journal = {\aap},
         year = 2008,
        month = dec,
       volume = {492},
       number = {1},
        pages = {277-287},
          doi = {10.1051/0004-6361:200810395},
archivePrefix = {arXiv},
       eprint = {0808.0270},
 primaryClass = {astro-ph},
       adsurl = {https://ui.adsabs.harvard.edu/abs/2008A&A...492..277B}
}

@INPROCEEDINGS{Castelli03,
       author = {{Castelli}, F. and {Kurucz}, R.~L.},
        title = "{New Grids of ATLAS9 Model Atmospheres}",
    booktitle = {Modelling of Stellar Atmospheres},
         year = 2003,
       editor = {{Piskunov}, N. and {Weiss}, W.~W. and {Gray}, D.~F.},
       series = {IAU Symposium},
       volume = {210},
        month = jan,
        pages = {A20},
          doi = {10.48550/arXiv.astro-ph/0405087},
archivePrefix = {arXiv},
       eprint = {astro-ph/0405087},
 primaryClass = {astro-ph},
       adsurl = {https://ui.adsabs.harvard.edu/abs/2003IAUS..210P.A20C}
}

@ARTICLE{Bressan12,
       author = {{Bressan}, Alessandro and {Marigo}, Paola and {Girardi}, L{\'e}o. and {Salasnich}, Bernardo and {Dal Cero}, Claudia and {Rubele}, Stefano and {Nanni}, Ambra},
        title = "{PARSEC: stellar tracks and isochrones with the PAdova and TRieste Stellar Evolution Code}",
      journal = {\mnras},
         year = 2012,
        month = nov,
       volume = {427},
       number = {1},
        pages = {127-145},
          doi = {10.1111/j.1365-2966.2012.21948.x},
archivePrefix = {arXiv},
       eprint = {1208.4498},
 primaryClass = {astro-ph.SR},
       adsurl = {https://ui.adsabs.harvard.edu/abs/2012MNRAS.427..127B}
}

@PROCEEDINGS{LB82,
        title = "{Landolt-B{\"o}rnstein: Numerical Data and Functional Relationships in Science and Technology - New Series `` Gruppe/Group 6 Astronomy and Astrophysics '' Volume 2 Schaifers/Voigt: Astronomy and Astrophysics / Astronomie und Astrophysik '' Stars and Star Clusters / Sterne und Sternhaufen}",
    booktitle = {Landolt-Bornstein: Group 6: Astronomy},
         year = 1982,
       editor = {{Aller}, L.~H. and {Appenzeller}, I. and {Baschek}, B. and {Duerbeck}, H.~W. and {Herczeg}, T. and {Lamla}, E. and {Meyer-Hofmeister}, E. and {Schmidt-Kaler}, T. and {Scholz}, M. and {Seggewiss}, W. and {Seitter}, W.~C. and {Weidemann}, V.},
       volume = {2},
        month = jan,
       adsurl = {https://ui.adsabs.harvard.edu/abs/1982lbg6.conf.....A}
}

@ARTICLE{Hog00,
       author = {{H{\o}g}, E. and {Fabricius}, C. and {Makarov}, V.~V. and {Urban}, S. and {Corbin}, T. and {Wycoff}, G. and {Bastian}, U. and {Schwekendiek}, P. and {Wicenec}, A.},
        title = "{The Tycho-2 catalogue of the 2.5 million brightest stars}",
      journal = {\aap},
         year = 2000,
        month = mar,
       volume = {355},
        pages = {L27-L30},
       adsurl = {https://ui.adsabs.harvard.edu/abs/2000A&A...355L..27H}
}

@ARTICLE{Norazman25,
       author = {{Norazman}, Azib and {Kennedy}, Grant M. and {Cody}, Ann Marie and {Giles}, Daniel and {Gill}, Samuel and {Kruse}, Ethan},
        title = "{A search for transiting exocomets in TESS sectors 1-26}",
      journal = {\mnras},
         year = 2025,
        month = sep,
       volume = {542},
       number = {2},
        pages = {1486-1508},
          doi = {10.1093/mnras/staf1298},
archivePrefix = {arXiv},
       eprint = {2508.04673},
 primaryClass = {astro-ph.EP},
       adsurl = {https://ui.adsabs.harvard.edu/abs/2025MNRAS.542.1486N}
}

@ARTICLE{Duchene13,
       author = {{Duch{\^e}ne}, Gaspard and {Kraus}, Adam},
        title = "{Stellar Multiplicity}",
      journal = {\araa},
         year = 2013,
        month = aug,
       volume = {51},
       number = {1},
        pages = {269-310},
          doi = {10.1146/annurev-astro-081710-102602},
archivePrefix = {arXiv},
       eprint = {1303.3028},
 primaryClass = {astro-ph.SR},
       adsurl = {https://ui.adsabs.harvard.edu/abs/2013ARA&A..51..269D}
}

@ARTICLE{Rodrigo24,
       author = {{Rodrigo}, Carlos and {Cruz}, Patricia and {Aguilar}, John F. and {Aller}, Alba and {Solano}, Enrique and {G{\'a}lvez-Ortiz}, Maria Cruz and {Jim{\'e}nez-Esteban}, Francisco and {Mas-Buitrago}, Pedro and {Bayo}, Amelia and {Cort{\'e}s-Contreras}, Miriam and {Murillo-Ojeda}, Raquel and {Bonoli}, Silvia and {Cenarro}, Javier and {Dupke}, Renato and {L{\'o}pez-Sanjuan}, Carlos and {Mar{\'\i}n-Franch}, Antonio and {de Oliveira}, Claudia Mendes and {Moles}, Mariano and {Taylor}, Keith and {Varela}, Jes{\'u}s and {Rami{\'o}}, H{\'e}ctor V{\'a}zquez},
        title = "{Photometric segregation of dwarf and giant FGK stars using the SVO Filter Profile Service and photometric tools}",
      journal = {\aap},
         year = 2024,
        month = sep,
       volume = {689},
          eid = {A93},
        pages = {A93},
          doi = {10.1051/0004-6361/202449998},
archivePrefix = {arXiv},
       eprint = {2406.03310},
 primaryClass = {astro-ph.SR},
       adsurl = {https://ui.adsabs.harvard.edu/abs/2024A&A...689A..93R}
}

@ARTICLE{Absil21,
       author = {{Absil}, O. and {Marion}, L. and {Ertel}, S. and {Defr{\`e}re}, D. and {Kennedy}, G.~M. and {Romagnolo}, A. and {Le Bouquin}, J. -B. and {Christiaens}, V. and {Milli}, J. and {Bonsor}, A. and {Olofsson}, J. and {Su}, K.~Y.~L. and {Augereau}, J. -C.},
        title = "{A near-infrared interferometric survey of debris-disk stars. VII. The hot-to-warm dust connection}",
      journal = {\aap},
         year = 2021,
        month = jul,
       volume = {651},
          eid = {A45},
        pages = {A45},
          doi = {10.1051/0004-6361/202140561},
archivePrefix = {arXiv},
       eprint = {2104.14216},
 primaryClass = {astro-ph.EP},
       adsurl = {https://ui.adsabs.harvard.edu/abs/2021A&A...651A..45A}
}

@ARTICLE{Merand05,
       author = {{M{\'e}rand}, A. and {Bord{\'e}}, P. and {Coud{\'e} du Foresto}, V.},
        title = "{A catalog of bright calibrator stars for 200-m baseline near-infrared stellar interferometry}",
      journal = {\aap},
         year = 2005,
        month = apr,
       volume = {433},
       number = {3},
        pages = {1155-1162},
          doi = {10.1051/0004-6361:20041323},
archivePrefix = {arXiv},
       eprint = {astro-ph/0412251},
 primaryClass = {astro-ph},
       adsurl = {https://ui.adsabs.harvard.edu/abs/2005A&A...433.1155M}
}

@ARTICLE{LeBouquin12,
       author = {{Le Bouquin}, J. -B. and {Absil}, O.},
        title = "{On the sensitivity of closure phases to faint companions in optical long baseline interferometry}",
      journal = {\aap},
         year = 2012,
        month = may,
       volume = {541},
          eid = {A89},
        pages = {A89},
          doi = {10.1051/0004-6361/201117891},
archivePrefix = {arXiv},
       eprint = {1204.3721},
 primaryClass = {astro-ph.IM},
       adsurl = {https://ui.adsabs.harvard.edu/abs/2012A&A...541A..89L}
}

@ARTICLE{Marion14,
       author = {{Marion}, L. and {Absil}, O. and {Ertel}, S. and {Le Bouquin}, J. -B. and {Augereau}, J. -C. and {Blind}, N. and {Defr{\`e}re}, D. and {Lebreton}, J. and {Milli}, J.},
        title = "{Searching for faint companions with VLTI/PIONIER. II. 92 main sequence stars from the Exozodi survey}",
      journal = {\aap},
         year = 2014,
        month = oct,
       volume = {570},
          eid = {A127},
        pages = {A127},
          doi = {10.1051/0004-6361/201424780},
archivePrefix = {arXiv},
       eprint = {1409.6105},
 primaryClass = {astro-ph.SR},
       adsurl = {https://ui.adsabs.harvard.edu/abs/2014A&A...570A.127M}
}

@ARTICLE{Ertel14,
       author = {{Ertel}, S. and {Absil}, O. and {Defr{\`e}re}, D. and {Le Bouquin}, J. -B. and {Augereau}, J. -C. and {Marion}, L. and {Blind}, N. and {Bonsor}, A. and {Bryden}, G. and {Lebreton}, J. and {Milli}, J.},
        title = "{A near-infrared interferometric survey of debris-disk stars. IV. An unbiased sample of 92 southern stars observed in H band with VLTI/PIONIER}",
      journal = {\aap},
         year = 2014,
        month = oct,
       volume = {570},
          eid = {A128},
        pages = {A128},
          doi = {10.1051/0004-6361/201424438},
archivePrefix = {arXiv},
       eprint = {1409.6143},
 primaryClass = {astro-ph.EP},
       adsurl = {https://ui.adsabs.harvard.edu/abs/2014A&A...570A.128E}
}

@ARTICLE{Kervella04,
       author = {{Kervella}, P. and {Th{\'e}venin}, F. and {Di Folco}, E. and {S{\'e}gransan}, D.},
        title = "{The angular sizes of dwarf stars and subgiants. Surface brightness relations calibrated by interferometry}",
      journal = {\aap},
         year = 2004,
        month = oct,
       volume = {426},
        pages = {297-307},
          doi = {10.1051/0004-6361:20035930},
archivePrefix = {arXiv},
       eprint = {astro-ph/0404180},
 primaryClass = {astro-ph},
       adsurl = {https://ui.adsabs.harvard.edu/abs/2004A&A...426..297K}
}

@ARTICLE{Moor25,
       author = {{Mo{\'o}r}, A. and {{\'A}brah{\'a}m}, P. and {K{\'o}sp{\'a}l}, {\'A}. and {Cataldi}, G. and {Hughes}, A.~M. and {Marino}, S. and {Kral}, Q. and {Milli}, J. and {Pawellek}, N.},
        title = "{Discovery of carbon monoxide emission from five debris disks around young A-type stars}",
      journal = {\aap},
         year = 2025,
        month = oct,
       volume = {703},
          eid = {A15},
        pages = {A15},
          doi = {10.1051/0004-6361/202554848},
archivePrefix = {arXiv},
       eprint = {2509.16104},
 primaryClass = {astro-ph.EP},
       adsurl = {https://ui.adsabs.harvard.edu/abs/2025A&A...703A..15M}
}

@ARTICLE{Kral20,
       author = {{Kral}, Quentin and {Matr{\`a}}, Luca and {Kennedy}, Grant M. and {Marino}, Sebastian and {Wyatt}, Mark C.},
        title = "{Survey of planetesimal belts with ALMA: gas detected around the Sun-like star HD 129590}",
      journal = {\mnras},
         year = 2020,
        month = sep,
       volume = {497},
       number = {3},
        pages = {2811-2830},
          doi = {10.1093/mnras/staa2038},
archivePrefix = {arXiv},
       eprint = {2005.05841},
 primaryClass = {astro-ph.EP},
       adsurl = {https://ui.adsabs.harvard.edu/abs/2020MNRAS.497.2811K}
}

@ARTICLE{Marino16,
       author = {{Marino}, S. and {Matr{\`a}}, L. and {Stark}, C. and {Wyatt}, M.~C. and {Casassus}, S. and {Kennedy}, G. and {Rodriguez}, D. and {Zuckerman}, B. and {Perez}, S. and {Dent}, W.~R.~F. and {Kuchner}, M. and {Hughes}, A.~M. and {Schneider}, G. and {Steele}, A. and {Roberge}, A. and {Donaldson}, J. and {Nesvold}, E.},
        title = "{Exocometary gas in the HD 181327 debris ring}",
      journal = {\mnras},
         year = 2016,
        month = aug,
       volume = {460},
       number = {3},
        pages = {2933-2944},
          doi = {10.1093/mnras/stw1216},
archivePrefix = {arXiv},
       eprint = {1605.05331},
 primaryClass = {astro-ph.EP},
       adsurl = {https://ui.adsabs.harvard.edu/abs/2016MNRAS.460.2933M}
}

@ARTICLE{Vrignaud2025,
       author = {{Vrignaud}, T. and {Lecavelier des Etangs}, A. and {Str{\o}m}, P.~A. and {Kiefer}, F.},
        title = "{Abundances of refractory ions in Beta Pictoris exocomets}",
      journal = {\aap},
         year = 2025,
        month = may,
       volume = {697},
          eid = {A21},
        pages = {A21},
          doi = {10.1051/0004-6361/202453568},
archivePrefix = {arXiv},
       eprint = {2503.17346},
 primaryClass = {astro-ph.EP},
       adsurl = {https://ui.adsabs.harvard.edu/abs/2025A&A...697A..21V}
}

@ARTICLE{Pavlenko22,
       author = {{Pavlenko}, Ya. and {Kulyk}, I. and {Shubina}, O. and {Vasylenko}, M. and {Dobrycheva}, D. and {Korsun}, P.},
        title = "{New exocomets of {\ensuremath{\beta}} Pic}",
      journal = {\aap},
         year = 2022,
        month = apr,
       volume = {660},
          eid = {A49},
        pages = {A49},
          doi = {10.1051/0004-6361/202142111},
archivePrefix = {arXiv},
       eprint = {2202.13373},
 primaryClass = {astro-ph.SR},
       adsurl = {https://ui.adsabs.harvard.edu/abs/2022A&A...660A..49P}
}

@ARTICLE{Dumond25,
       author = {{Dumond}, P. and {Lecavelier des Etangs}, A. and {Kiefer}, F. and {H{\'e}brard}, G. and {Caill{\'e}}, V.},
        title = "{Search for exocomet transits in Kepler light curves: Ten new transits identified}",
      journal = {\aap},
         year = 2025,
        month = dec,
       volume = {704},
          eid = {A191},
        pages = {A191},
          doi = {10.1051/0004-6361/202556033},
archivePrefix = {arXiv},
       eprint = {2510.14687},
 primaryClass = {astro-ph.EP},
       adsurl = {https://ui.adsabs.harvard.edu/abs/2025A&A...704A.191D}
}

@ARTICLE{Howarth86,
       author = {{Howarth}, Ian D. and {Phillips}, A. Paul},
        title = "{The ultraviolet spectrum and interstellar environment of HD 50896.}",
      journal = {\mnras},
         year = 1986,
        month = oct,
       volume = {222},
        pages = {809-852},
          doi = {10.1093/mnras/222.4.809},
       adsurl = {https://ui.adsabs.harvard.edu/abs/1986MNRAS.222..809H}
}

@INPROCEEDINGS{Currie14,
       author = {{Currie}, M.~J. and {Berry}, D.~S. and {Jenness}, T. and {Gibb}, A.~G. and {Bell}, G.~S. and {Draper}, P.~W.},
        title = "{Starlink Software in 2013}",
    booktitle = {Astronomical Data Analysis Software and Systems XXIII},
         year = 2014,
       editor = {{Manset}, N. and {Forshay}, P.},
       series = {Astronomical Society of the Pacific Conference Series},
       volume = {485},
        month = may,
        pages = {391},
       adsurl = {https://ui.adsabs.harvard.edu/abs/2014ASPC..485..391C}
}

\clearpage
\onecolumn
\begin{appendix}
\section{Data}
\label{app:data}

\begin{table*}[ht]
    \caption{Summary of results from the non-photospheric Ca {\sc ii} K components.}
    \label{tab:alldata}
    \centering
    \begin{tabular}{c c c c| c c c}
    \hline\hline
    \noalign{\smallskip}
    Star & Date (D/M/Y) & EW (m$\AA$) & N (cm$^{-2}$) & Date (D/M/Y) & EW (m$\AA$) & N (cm$^{-2}$)\\
    \noalign{\smallskip}
    \hline
    \noalign{\smallskip}
    HD~36546 & 06/03/2017 (1) & $30.7\pm3.1$ & 3.29$\cdot10^{11}$  & 30/11/2022 (2)   & $31.0\pm0.8$ & 3.32$\cdot10^{11}$ \\
             & 07/03/2017 (2) & $30.4\pm1.7$ & 3.26$\cdot10^{11}$  & 03/12/2022 (2)   & $30.6\pm0.6$ & 3.28$\cdot10^{11}$ \\
             & 08/03/2017 (1) & $28.4\pm2.0$ & 3.04$\cdot10^{11}$  & 24/12/2022 (2)   & $29.9\pm1.3$ & 3.20$\cdot10^{11}$ \\
             & 09/03/2017 (1) & $30.7\pm1.7$ & 3.29$\cdot10^{11}$  & 13/01/2023 (4)   & $30.3\pm1.0$ & 3.24$\cdot10^{11}$   \\
             & 10/03/2017 (1) & $30.0\pm1.6$ & 3.21$\cdot10^{11}$  & 14/01/2023 (8)   & $29.2\pm0.7$ & 3.13$\cdot10^{11}$  \\
             & 12/03/2017 (1) & $31.6\pm1.6$ & 3.38$\cdot10^{11}$  & 15/01/2023 (2)   & $28.9\pm0.9$ & 3.09$\cdot10^{11}$ \\
             & 01/04/2017 (1) & $29.2\pm1.3$ & 3.13$\cdot10^{11}$  & 26/01/2023 (2)   & $28.4\pm0.8$ & 3.04$\cdot10^{11}$ \\
             & 15/11/2022 (2) & $31.0\pm0.9$ & 3.32$\cdot10^{11}$  & {\bf Total (32)} & $30.0\pm0.3$ & 3.21$\cdot10^{11}$ \\
     \noalign{\smallskip}
     \hline
     \noalign{\smallskip}
    HD~42111 & 04/09/2015 (1) & $327\pm8$  & 3.50$\cdot10^{12}$  & 04/04/2017 (1) & $322\pm5$       & 3.45$\cdot10^{12}$  \\
             & 05/09/2015 (2) & $337\pm5$  & 3.61$\cdot10^{12}$  & 05/04/2017 (1) & $316\pm5$       & 3.38$\cdot10^{12}$ \\
             & 21/12/2015 (1) & $322\pm5$  & 3.45$\cdot10^{12}$  & 06/04/2017 (1) & $349\pm5$       & 3.74$\cdot10^{12}$ \\
             & 23/12/2015 (1) & $323\pm4$  & 3.46$\cdot10^{12}$  & 08/04/2017 (1) & $335\pm7$       & 3.59$\cdot10^{12}$ \\
             & 24/12/2015 (1) & $316\pm4$  & 3.38$\cdot10^{12}$  & 09/04/2017 (1) & $342\pm6$       & 3.66$\cdot10^{12}$ \\
             & 28/01/2016 (2) & $320\pm5$  & 3.43$\cdot10^{12}$  & 03/11/2022 (1) & $379\pm3$       & 4.06$\cdot10^{12}$ \\
             & 31/01/2016 (1) & $274\pm10$ & 2.93$\cdot10^{12}$  & 14/11/2022 (1) & $373\pm4$       & 3.99$\cdot10^{12}$ \\
             & 03/03/2016 (1) & $304\pm5$  & 3.26$\cdot10^{12}$  & 15/11/2022 (1) & $384\pm4$       & 4.11$\cdot10^{12}$ \\
             & 04/03/2016 (1) & $307\pm5$  & 3.29$\cdot10^{12}$  & 01/12/2022 (1) & $377\pm3$       & 4.04$\cdot10^{12}$ \\
             & 05/03/2016 (2) & $304\pm4$  & 3.26$\cdot10^{12}$  & 03/12/2022 (1) & $356\pm3$       & 3.81$\cdot10^{12}$ \\
             & 06/03/2016 (1) & $307\pm5$  & 3.29$\cdot10^{12}$  & 14/12/2022 (1) & $359\pm5$       & 3.84$\cdot10^{12}$ \\
             & 08/03/2017 (1) & $309\pm5$  & 3.31$\cdot10^{12}$  & 01/01/2023 (1) & $368\pm4$       & 3.94$\cdot10^{12}$ \\
             & 09/03/2017 (1) & $310\pm5$  & 3.32$\cdot10^{12}$  & 13/01/2023 (2) & $345\pm3$       & 3.69$\cdot10^{12}$ \\
             & 11/03/2017 (2) & $250\pm5$  & 2.68$\cdot10^{12}$  & 14/01/2023 (1) & $335\pm4$       & 3.59$\cdot10^{12}$ \\
             & 12/03/2017 (1) & $314\pm5$  & 3.365$\cdot10^{12}$ & 15/01/2023 (3) & $327\pm3$       & 3.50$\cdot10^{12}$ \\
             & 13/03/2017 (1) & $311\pm5$  & 3.33$\cdot10^{12}$  & 17/01/2023 (1) & $351\pm4$       & 3.76$\cdot10^{12}$ \\
             & 29/03/2017 (1) & $313\pm4$  & 3.35$\cdot10^{12}$  & 24/01/2023 (1) & $349\pm5$       & 3.74$\cdot10^{12}$ \\
             & 30/03/2017 (1) & $328\pm4$  & 3.51$\cdot10^{12}$  & 28/01/2023 (1) & $349\pm4$       & 3.74$\cdot10^{12}$ \\
             & 02/04/2017 (1) & $327\pm5$  & 3.50$\cdot10^{12}$  & {\bf Total (45)} & $339\pm1$  & 3.63$\cdot10^{12}$ \\
             & 03/04/2017 (1) & $303\pm4$  & 3.24$\cdot10^{12}$  & \\ 
     \noalign{\smallskip}
     \hline
     \noalign{\smallskip}
     HD~85905& 15/12/2012 (2) & $110.1\pm1.8$  & 11.79$\cdot10^{11}$  & 05/04/2017 (1) & $71.0\pm1.9$      & 7.60$\cdot10^{11}$ \\
             & 21/12/2015 (2) & $85.5\pm4.1$   &  9.16$\cdot10^{11}$  & 06/04/2017 (1) & $81.7\pm2.3$      & 8.75$\cdot10^{11}$ \\
             & 23/12/2015 (3) & $77.7\pm3.0$   &  8.32$\cdot10^{11}$  & 07/04/2017 (1) & $82.2\pm3.0$      & 8.80$\cdot10^{11}$ \\
             & 24/12/2015 (3) & $87.3\pm3.2$   &  9.35$\cdot10^{11}$  & 08/04/2017 (2) & $83.2\pm2.4$      & 8.91$\cdot10^{11}$ \\    
             & 27/01/2016 (1) & $90.3\pm2.9$   &  9.67$\cdot10^{11}$  & 09/04/2017 (1) & $84.2\pm2.7$      & 9.02$\cdot10^{11}$ \\
             & 29/01/2016 (2) & $84.7\pm3.4$   &  9.07$\cdot10^{11}$  & 15/11/2022 (1) & $62.8\pm3.5$      & 6.72$\cdot10^{11}$ \\
             & 03/03/2016 (2) & $84.7\pm4.1$   &  9.07$\cdot10^{11}$  & 18/11/2022 (1) & $52.7\pm4.0$      & 5.64$\cdot10^{11}$ \\
             & 04/03/2016 (4) & $83.8\pm2.9$   &  8.97$\cdot10^{11}$  & 01/12/2022 (1) & $47.0\pm3.8$      & 5.03$\cdot10^{11}$ \\
             & 05/03/2016 (3) & $81.4\pm2.6$   &  8.72$\cdot10^{11}$  & 03/12/2022 (1) & $47.7\pm3.0$      & 5.11$\cdot10^{11}$ \\
             & 06/03/2016 (4) & $87.0\pm2.7$   &  9.32$\cdot10^{11}$  & 15/12/2022 (1) & $59.1\pm4.2$      & 6.33$\cdot10^{11}$ \\            
             & 07/03/2016 (1) & $76.3\pm3.8$   &  8.17$\cdot10^{11}$  & 14/01/2023 (3) & $47.3\pm2.4$      & 5.06$\cdot10^{11}$ \\
             & 11/03/2017 (2) & $74.3\pm4.3$   &  7.96$\cdot10^{11}$  & 15/01/2023 (4) & $50.6\pm2.0$      & 5.42$\cdot10^{11}$ \\  
             & 29/03/2017 (1) & $74.9\pm2.8$   &  8.02$\cdot10^{11}$  & 14/02/2023 (2) & $54.2\pm2.3$      & 5.80$\cdot10^{11}$ \\  
             & 31/03/2017 (1) & $78.7\pm3.0$   &  8.43$\cdot10^{11}$  & 12/03/2023 (2) & $37.2\pm4.7$      & 3.98$\cdot10^{11}$ \\
             & 01/04/2017 (1) & $69.9\pm2.3$   &  7.48$\cdot10^{11}$  & 24/04/2025 (1) & $76.0\pm6.8$      & 8.14$\cdot10^{11}$ \\
             & 02/04/2017 (3) & $76.7\pm1.8$   &  8.21$\cdot10^{11}$  & 28/04/2025 (5) & $68.5\pm2.8$      & 7.33$\cdot10^{11}$ \\ 
             & 03/04/2017 (2) & $76.0\pm1.9$   &  8.14$\cdot10^{11}$  & {\bf Total (66)} & $70.7\pm0.7$ & 7.57$\cdot10^{11}$ \\  
             & 04/04/2017 (1) & $77.0\pm3.6$   &  8.25$\cdot10^{11}$  & \\              
   \end{tabular}
   \tablefoot{Columns are: object, date of observation (number of spectra taken on that date), equivalent widths and column densities from the mean weighted spectra of each date. {\bf Total} refers to the values obtained from the weighted average spectrum built using all the available spectra for each object; these values appear as EW$_{\rm average}$ and N$_{\rm average}$ in Table \ref{tab:paramsfromews}.}
\end{table*}

\begin{figure*}[!ht]
    \centering
    \includegraphics[width=0.85\linewidth]{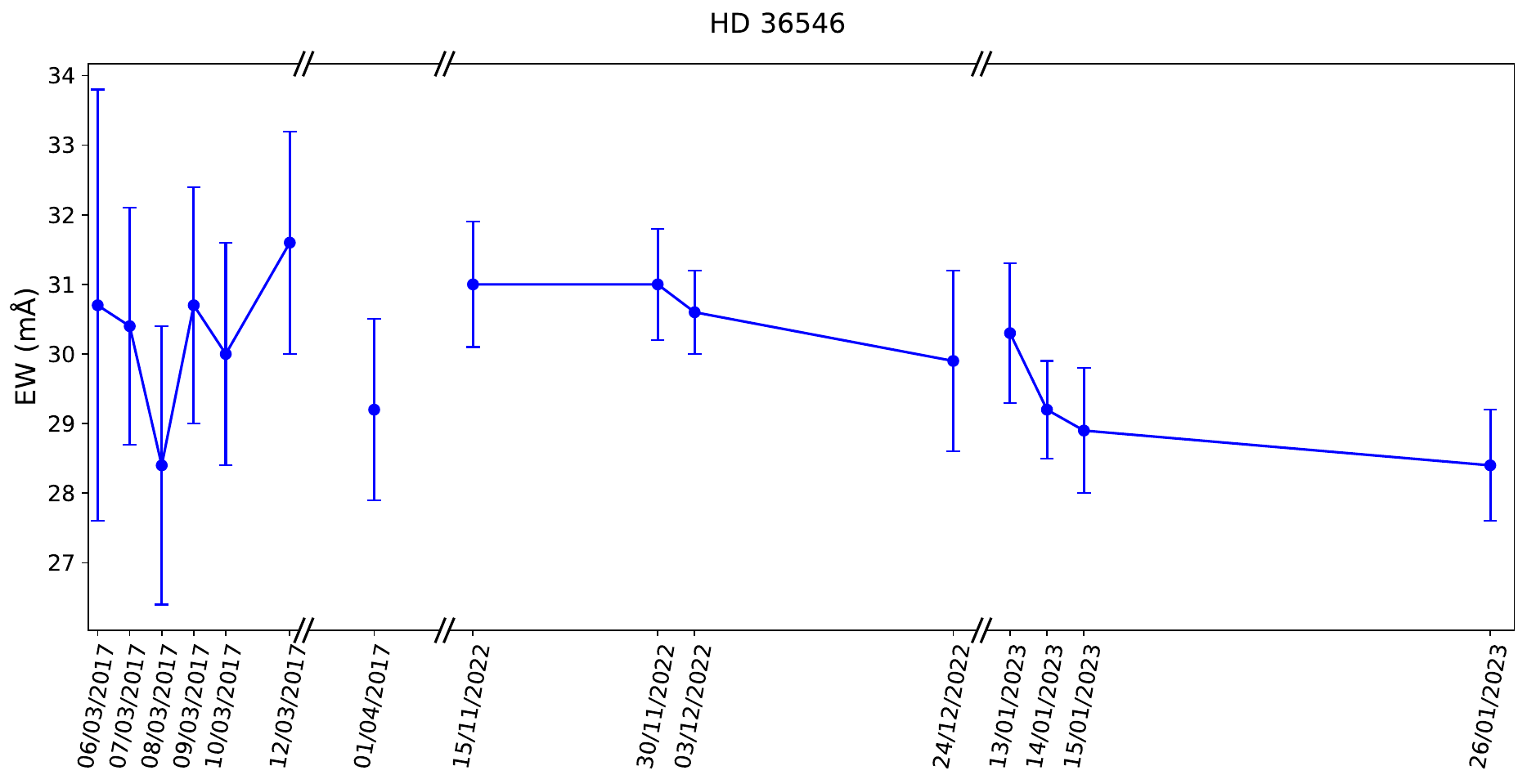}
    \includegraphics[width=0.85\linewidth]{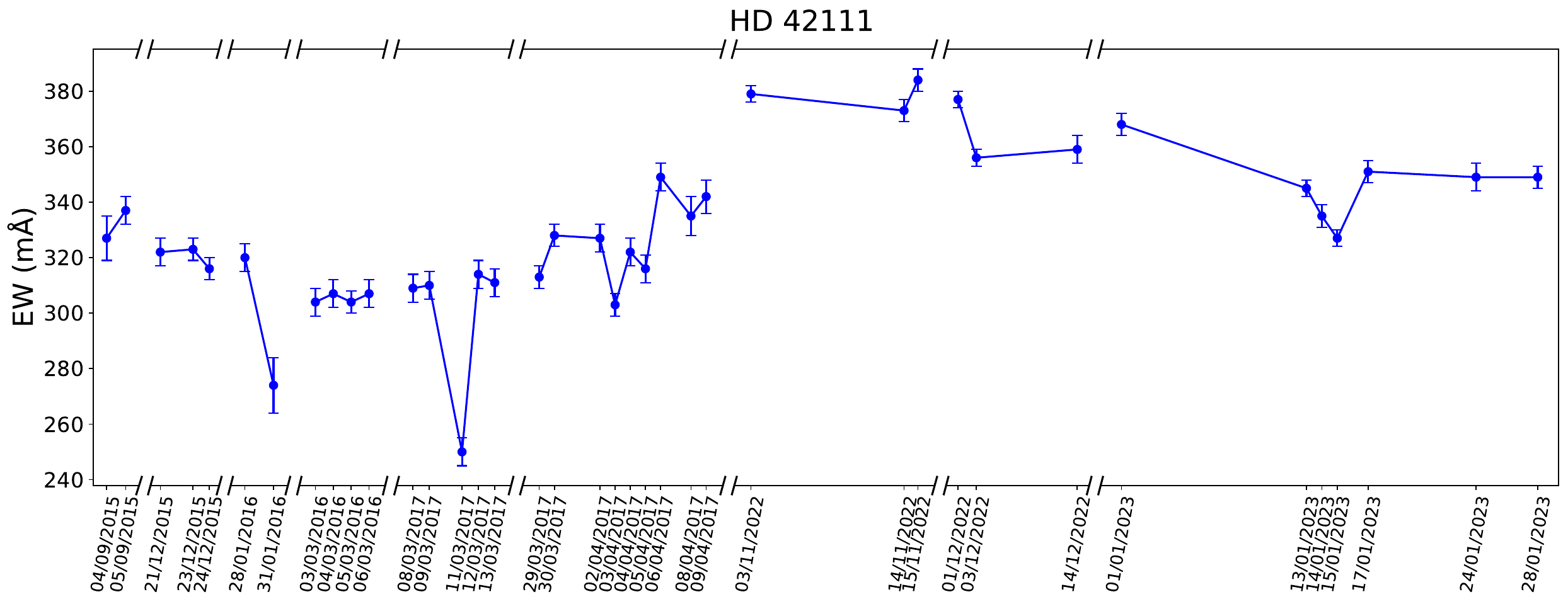}
    \includegraphics[width=0.85\linewidth]{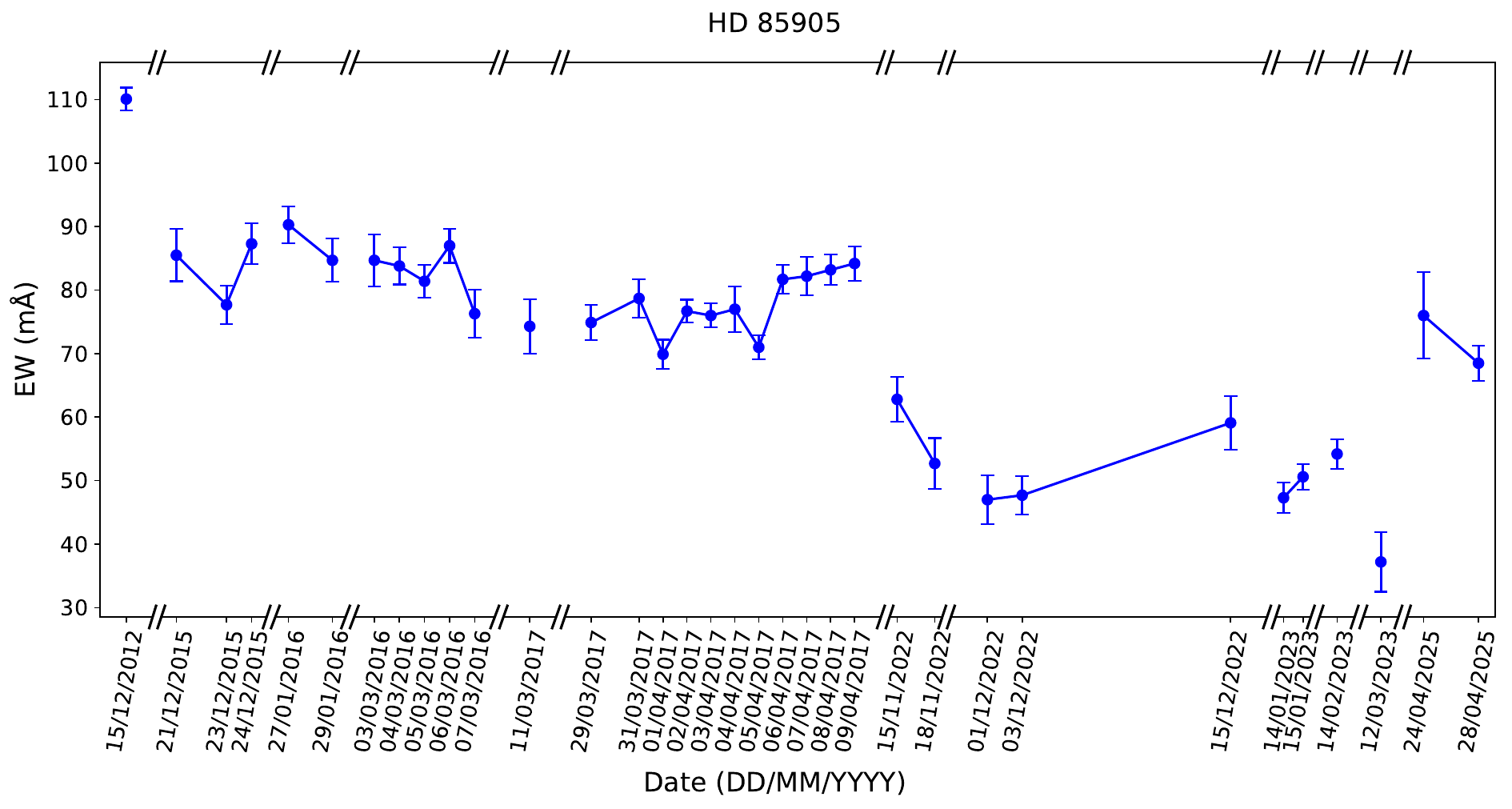}
    \caption{Variation of the equivalent widths of the circumstellar 
    absorptions in the Ca {\sc ii} K line with time. The data displayed
    in these panels correspond to those in Table \ref{tab:alldata}.}
    \label{fig:EW_time_ev}
\end{figure*}

\begin{figure*}[!ht]
    \centering
    \includegraphics[width=1.0\linewidth]{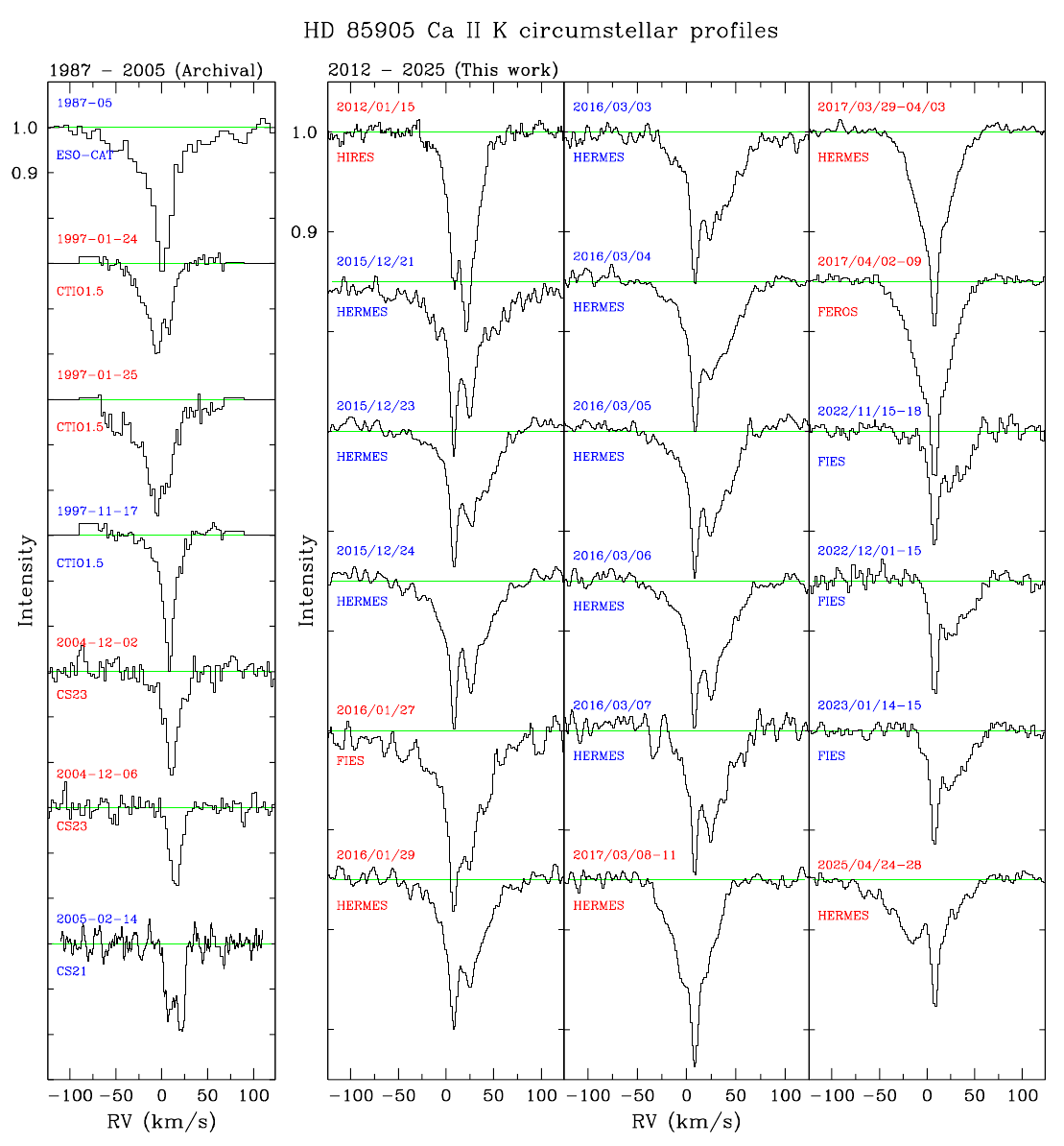}
    \caption{circumstellar components of the Ca {\sc ii} K line for HD~85905 in multiple epochs. The panel on the left shows spectra scanned from papers (1987/05, 1997/01 and 1997/11), and archival data (2004 and 2005) obtained with the instruments CS21 and CS23 on the Harlan J. Smith 2.7m Telescope at McDonald Observatory (see text for details and references about all these observations). The three panels on the right show the circumstellar profiles of observations obtained in the period 2012--2025 which have been used to derive the results shown in Table \ref{tab:alldata}. Red and blue labels alternate to separate observations of different campaigns, this helps to make comparisons of the differences between profiles observed in runs separated months or years. In addition to the dates, the spectrographs used are specified in the labels. The data in the panel on the left are shown just for comparison with later spectra, and have not been used in the analysis carried out in this paper.}
    \label{HD85905all}
\end{figure*}

\clearpage
\twocolumn
\section{HD~85905: photometry, SED fitting and HR diagrams}
\label{app:photometryandhr}

\subsection{Photometry}

The Virtual Observatory SED Analyzer \citep[VOSA;][]{Bayo08} was used to 
collect the photometry across a wide spectral range and fit the SED of HD~85905. Table \ref{tab:photometry} shows the optical, near-, and mid-IR photometric data converted into fluxes, with the corresponding uncertainties, collected by VOSA. Synthetic photometry, performed over a STIS/HST spectrum\footnote{The STIS spectrum was obtained within the program corresponding to HST proposal 9431, cycle 11, PI Alain Lecavelier des Etangs.}, was generated using the SpecPhot\footnote{SpecPhot is available at \url{https://svo2.cab.inta-csic.es/theory/specphot/index.php}.} application \citep{Rodrigo24} and it was also added to the 
available photometric data.

Regarding the optical photometry, which is important to estimate the brightness of the object, SIMBAD lists the following values:
$B\!=\!6.270\pm0.014$, $V\!=\!6.213\pm0.010$, and gives as a reference
the Tycho-2 catalogue \citep{Hog00}, which does not list the Johnson magnitudes, but the Tycho $B_{\rm T}\!=\!6.291\pm0.014$, $V_{\rm T}\!=\!6.221\pm0.010$ magnitudes
instead; these can be transformed into the Johnson-Cousins system using the expressions\footnote{See eqns. 2.2.1 and 2.2.2 in \url{https://cdsarc.u-strasbg.fr/ftp/cats/more/HIP/cdroms/docs/vol1//sect2_02.pdf}}:

\begin{equation}
\begin{array}{lcl}
V      &\!=\!& V_{\rm T} - 0.090\cdot(B_{\rm T}\!-\!V_{\rm T}), \\
 B-V   &\!=\!& 0.850\cdot(B_{\rm T}\!-\!V_{\rm T})
\end{array}
\end{equation}

\noindent which yield $V\!=\!6.215\pm0.011$, $B\!=\!6.274\pm0.011$,
$B-V\!=\!0.059\pm0.015$.

On the other hand, the VizieR catalogue II/168/ubvmeans\footnote{\url{https://vizier.cds.unistra.fr/viz-bin/VizieR?-source=II/168}}, which lists the "Johnson magnitudes between 5000 and 6000 \AA" (sic), gives $V\!=\!6.233\pm0.005$, $B-V\!=\!0.040\pm0.005$.

An average of the two sets of values has been adopted as weighted means with inverse-variance weights, namely,  $B\!=\!6.273\pm0.005$, $V\!=\!6.230\pm0.005$, $B-V\!=\!0.043\pm0.006$.

\begin{table}[ht]
    \caption{HD~85905: Photometry}
    \label{tab:photometry}
    \centering
    \begin{tabular}{r c l}
    \hline\hline
    \noalign{\smallskip}
    \multicolumn{1}{c}{Wavelength} &  \multicolumn{1}{c}{Flux}       &    Filter \\   
    \multicolumn{1}{c}{(\AA)}      & (erg cm$^{-2}$ s$^{-1}$ \AA$^{-1}$) &         \\
    \noalign{\smallskip}
    \hline
    \noalign{\smallskip}
    3482.6 &  2.567 $\pm$ 0.017$ [\cdot 10^{-11}$] &  Str\"omgren $u$  \\ 
    3551.1 &  1.459 $\pm$ 0.036$ [\cdot 10^{-11}$] &  Johnson $U$  \\ 
    3608.0 &  1.532 $\pm$ 0.037$ [\cdot 10^{-11}$] & SDSS $u$  \\ 
    4124.3 &  1.959 $\pm$ 0.006$ [\cdot 10^{-11}$] &  Str\"omgren $v$  \\ 
    4280.0 &  1.982 $\pm$ 0.026$ [\cdot 10^{-11}$] &  Tycho $B$  \\ 
    4369.5 &  2.076 $\pm$ 0.015$ [\cdot 10^{-11}$] &  Johnson $B$  \\ 
    4369.5 &  2.082 $\pm$ 0.001$ [\cdot 10^{-11}$] &  Johnson $B$  \\ 
    4369.5 &  2.064 $\pm$ 0.018$ [\cdot 10^{-11}$] &  Johnson $B$  \\ 
    4666.9 &  1.809 $\pm$ 0.004$ [\cdot 10^{-11}$] &  Str\"omgren $b$  \\ 
    4671.8 &  1.791 $\pm$ 0.008$ [\cdot 10^{-11}$] &  SDSS $g$  \\ 
    4901.7 &  1.382 $\pm$ 0.001$ [\cdot 10^{-11}$] &  Hipparcos $H$p  \\ 
    5035.8 &  1.315 $\pm$ 0.004$ [\cdot 10^{-11}$] &  Gaia $G_{\rm bp}$  \\ 
    5340.0 &  1.298 $\pm$ 0.012$ [\cdot 10^{-11}$] &  Tycho $V$  \\ 
    5464.7 &  1.169 $\pm$ 0.002$ [\cdot 10^{-11}$] &  Str\"omgren $y$  \\ 
    5467.6 &  1.168 $\pm$ 0.005$ [\cdot 10^{-11}$] &  Johnson $V$  \\ 
    5467.6 &  1.151 $\pm$ 0.005$ [\cdot 10^{-11}$] &  Johnson $V$  \\ 
    5467.6 &  1.180 $\pm$ 0.004$ [\cdot 10^{-11}$] &  Johnson $V$  \\ 
    5809.3 &  9.663 $\pm$ 0.020$ [\cdot 10^{-12}$] &  HST ACS\_WFC/F606W \\ 
    5822.4 &  8.233 $\pm$ 0.021$ [\cdot 10^{-12}$] &  Gaia $G$  \\ 
    6141.1 &  8.439 $\pm$ 0.417$ [\cdot 10^{-12}$] &  SDSS $r$  \\ 
    6141.1 &  8.577 $\pm$ 0.025$ [\cdot 10^{-12}$] &  SDSS $r$  \\ 
    6695.8 &  7.441 $\pm$ 0.022$ [\cdot 10^{-12}$] &  Johnson $R$  \\ 
    7457.9 &  5.019 $\pm$ 0.016$ [\cdot 10^{-12}$] &  SDSS $i$  \\ 
    7620.0 &  4.487 $\pm$ 0.016$ [\cdot 10^{-12}$] &  Gaia $G_{\rm rp}$  \\ 
    7973.4 &  4.135 $\pm$ 0.014$ [\cdot 10^{-12}$] &  HST ACS\_WFC/F814W \\ 
    8568.9 &  4.164 $\pm$ 0.015$ [\cdot 10^{-12}$] &  Johnson $I$  \\ 
    8922.8 &  3.173 $\pm$ 0.014$ [\cdot 10^{-12}$] &  SDSS $z$  \\ 
    9613.6 &  2.710 $\pm$ 0.018$ [\cdot 10^{-12}$] &  PAN-STARRS $y$ \\ 
    12350.0 & 1.193 $\pm$ 0.022$ [\cdot 10^{-12}$] &  2MASS $J$  \\ 
    16620.0 & 4.367 $\pm$ 0.011$ [\cdot 10^{-13}$] &  2MASS $H$  \\ 
    21590.0 & 1.651 $\pm$ 0.032$ [\cdot 10^{-13}$] &  2MASS $K$s  \\ 
    33526.0 & 4.877 $\pm$ 0.277$ [\cdot 10^{-14}$] &  WISE W1  \\ 
    33526.0 & 5.763 $\pm$ 0.191$ [\cdot 10^{-14}$] &  WISE W1  \\ 
    33526.0 & 3.289 $\pm$ 0.342$ [\cdot 10^{-14}$] &  WISE W1  \\ 
    46028.0 & 1.380 $\pm$ 0.038$ [\cdot 10^{-14}$] &  WISE W2  \\ 
    46028.0 & 1.559 $\pm$ 0.027$ [\cdot 10^{-14}$] &  WISE W2  \\ 
    46028.0 & 1.079 $\pm$ 0.046$ [\cdot 10^{-14}$] &  WISE W2  \\ 
    82281.5 & 9.879 $\pm$ 0.192$ [\cdot 10^{-16}$] &  AKARI IRC/S9W  \\ 
    115608.0 & 2.565 $\pm$ 0.035$ [\cdot 10^{-16}$] &  WISE W3  \\ 
    220883.0 & 2.000 $\pm$ 0.079$ [\cdot 10^{-17}$] &  WISE W4  \\ 
    \noalign{\smallskip}
   \hline        
   \end{tabular}
\end{table}

\subsection{SED fitting}

VOSA was used to fit the SED of HD~85905. Among the collection of theoretical spectra available in that tool, Kurucz NEWODF models with no overshooting and $\alpha\!=\!0.0$ \citep{Castelli03} were selected. The value A$_V\!=\!0.237$ (\textit{Gaia} DR3) was used, and solar metallicity was assumed. The best fit was achieved for $T_{\rm eff}\!=\!9000$ K and $\log g\!=\!4.00$, a temperature very close to that --9040 K-- found by \cite{Rebollido20}. Note that, as it was mentioned in Sect. \ref{sec:parameters}, the analysis showed no indication that two models, potentially representative of the two components of the binary, were needed for the photometric fit, as it happened with the spectroscopic analysis. This fact was an important piece of information suggesting that the binary could be composed of an almost-equal pair of stars.

In addition to the PIONIER observation, binarity was also supported by the inconsistency
between the luminosity provided by the photometric fit and the brightness derived from
the optical photometry. Correcting for reddening with A$_V\!=\!0.237$ (\textit{Gaia} DR3), the intrinsic values of the $V$ magnitude and the $B\!-\!V$ colour are $V_0\!=\!5.993$, $(B\!-\!V)_0\!=\!-0.033$, and using the distance modulus, the absolute magnitude is $M_V\!=\!-0.323$: this value does not correspond to that of a single star of spectral type A1 IV, the one listed in the SIMBAD database. According to the Schmidt-K\"aler compilation of standard data for stars \citep[][eds]{LB82}  --given the difficulty of finding these tables online, a excerpt with the relevant data for luminosity classes V and III is given in Table \ref{tab:schmidt-kaler}--, an object with $T_{\rm eff}\!\simeq9000$ K in the main sequence would have typical $M_V$ and $L_{\rm bol}$ of $\sim\!+1.2$ and $\sim\!30$ $L_\odot$, whereas the values for a giant with the same temperature would be $\sim\!+0.3$ and $\sim\!65$ $L_\odot$, respectively. Both luminosities are well below that inferred for HD~85905 from the SED analysis. On the other hand, the value of the $(B\!-\!V)_0\!=\!-0.033$ is slightly bluer than that corresponding to a $\sim\!9000$ K object. In the analysis carried out in Sect. \ref{sec:parameters} the temperature $T\!=\!9040$ K from the spectroscopic analysis, was adopted as the reference to carry out the derivation of the parameters of both components of the binary described in Sect. \ref{sec:parameters}.

\begin{table}[ht]
    \caption{Data for standard stars}
    \label{tab:schmidt-kaler}
    \centering
    \setlength{\tabcolsep}{9.0 pt}
    \begin{tabular}{l r c r}
    \hline\hline
    \noalign{\smallskip}
    Sp & \multicolumn{1}{c}{$T_{\rm eff}$ [K]} &  $M_V$  & $L/L_\odot$   \\ 
    \noalign{\smallskip}
    \hline
    \noalign{\smallskip}
        \multicolumn{4}{l}{Luminosity class V} \\
    \noalign{\smallskip}
    \hline
    \noalign{\smallskip}
    B8 & $11\,900$ & $-0.2$ & 180 \\ 
    B9 & $10\,500$ & $+0.2$ & 95  \\
    A0 & 9520      & $+0.6$ & 54  \\
    A1 & 9230      & $+1.0$ & 35  \\
    A2 & 8970      & $+1.3$ & 26  \\
    A3 & 8720      & $+1.5$ & 21  \\
    \noalign{\smallskip}
    \hline
    \noalign{\smallskip}
    \multicolumn{4}{l}{Luminosity class III} \\
    \noalign{\smallskip}
   \hline   
   \noalign{\smallskip}
   B8 & $12\,400$ & $-1.2$ & 460  \\
   B9 & $11\,000$ & $-0.6$ & 240  \\
   A0 & $10\,100$ & $+0.0$ & 106  \\
   A1 & $9480$    & $+0.2$ & 78   \\
   A2 & $9000$    & $+0.3$ & 65   \\
   A3 & $8600$    & $+0.5$ & 53   \\
   \noalign{\smallskip}
   \hline
   \end{tabular}
\end{table}   

\begin{figure*}[!ht]
    \centering
    \includegraphics[width=1.0\linewidth]{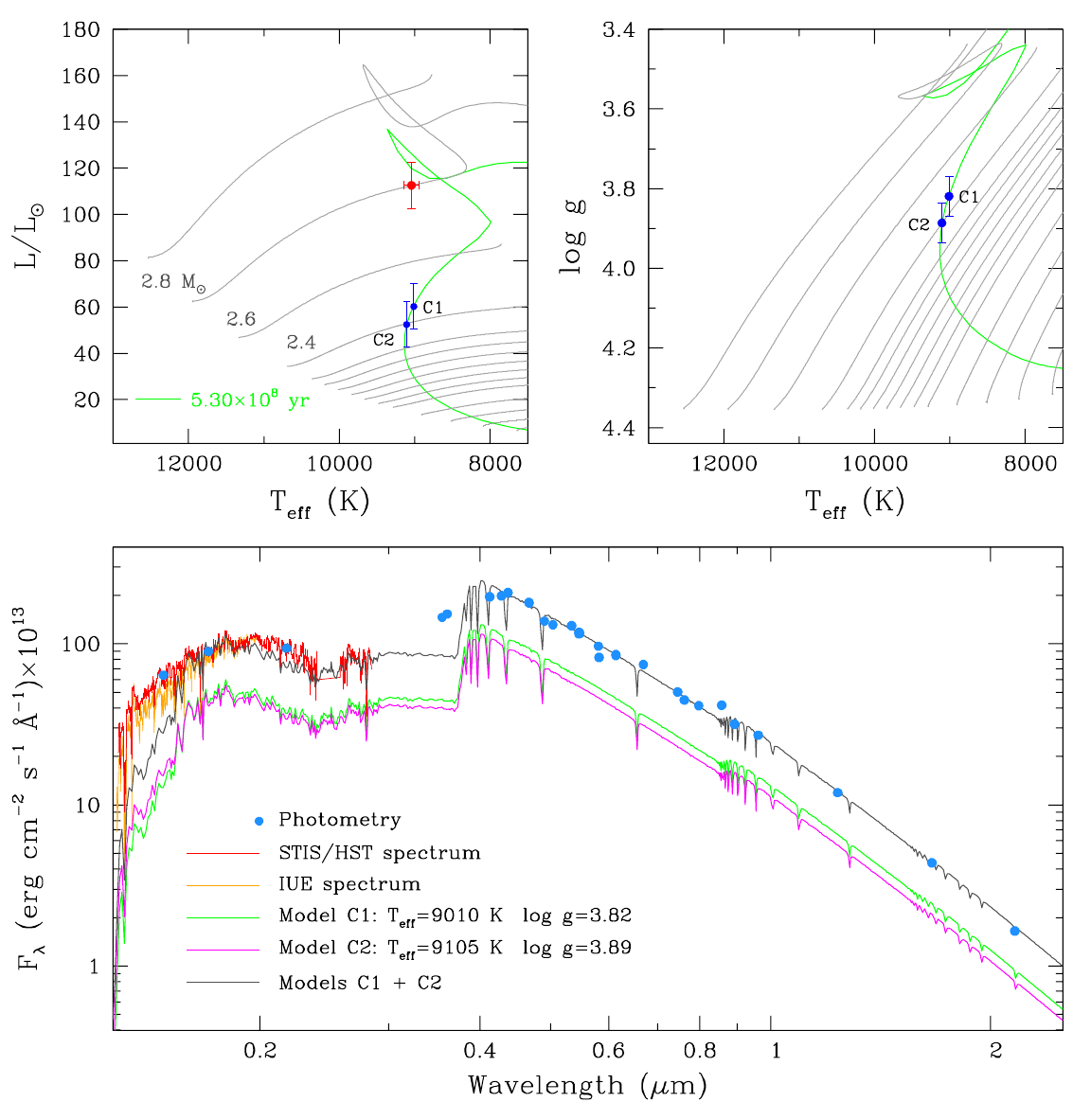}
    \caption{\textit{Upper panels}: HR diagrams. The luminosity 112.6 $L_\odot$,  plotted as a red dot on the $L_*/L_\odot$ - $T_{\rm eff}$ HR diagram, was derived by integrating a single-temperature (1T) dereddened model with $T_{\rm eff}\!=\!9040$ K, which also fits the SED giving a value of the total flux, $F_*$. The \textit{Gaia} DR3 parallax and the expression $L_*\!=\!4\,\pi\,d^2\,F_*$ 
    allowed us to compute the luminosity. The two blue dots in both diagrams correspond to parameters consistent with the binary scenario where the SED is reproduced by models for two coeval stars with parameters ($T_{\rm eff}, \log g$) (9010 K, 3.82) (9105 K, 3.89), for components 1 and 2 (labelled C1, C2), respectively, whose luminosities add up to the total luminosity of 112.6 
    $L_\odot$. PARSEC 2.1s tracks and isochrones were used in this analysis. \textit{Lower panels}: Observed photometry (blue dots), STIS/HST and IUE spectra of HD~85905 (red and orange lines, respectively), and the Kurucz photospheric models for components C1 and C2 built with the parameters shown in the legend, reddened with A$_V$=0.237. The models fulfil the observational constraint of a contrast in the flux in $H$-band of $0.855\pm0.076$ from the PIONIER data (see text for details).}
    \label{fig:sedandevolution}
\end{figure*}

\end{appendix}

\end{document}